\documentclass[journal]{IEEEtran}
\usepackage{algorithm}
\usepackage{algorithmic}
\usepackage{cite}
\usepackage{bbm}
\usepackage{amsmath}
\usepackage{amsfonts,amssymb}
\usepackage{graphicx}
\usepackage{colortbl}
\allowdisplaybreaks[4]
\usepackage{enumitem}
\usepackage{booktabs}
\usepackage{xcolor,colortbl}

 \usepackage{bm}
 \usepackage{multirow}
\allowdisplaybreaks[4]
\usepackage{amsthm}
\usepackage{amssymb}
\usepackage{pgfplots}
\usepackage[a]{esvect}

\usepackage{comment}
\definecolor{myblue}{RGB}{105,89,205}
\definecolor{myorange}{RGB}{238,92,66}
\definecolor{mygrey}{RGB}{205,201,201}

\begin{document}

\title{On the Optimality of Voltage Unbalance Attenuation by Inverters}
\author{
	Yifei~Guo,~\IEEEmembership{Member,~IEEE,}
	Bikash C. Pal,~\IEEEmembership{Fellow,~IEEE,}
	Rabih A. Jabr,~\IEEEmembership{Fellow,~IEEE}
%\thanks{This work was supported by the EPSRC-NSFC Programme on Sustainable Energy Supply. (\emph{Corresponding author: Yifei Guo})}
\thanks{Yifei Guo and Bikash C. Pal are with the Electrical and Electronic
Engineering Department, Imperial College London, London SW7 2AZ, U.K. (e-mail: yifei.guo@imperial.ac.uk; b.pal@imperial.ac.uk)}
\thanks{Rabih A. Jabr is with the Department of Electrical and Computer
Engineering, American University of Beirut, Beirut 1107 2020, Lebanon
(e-mail: rabih.jabr@aub.edu.lb)}
}
\markboth{}%This work has been submitted to the IEEE for possible publication. Copyright may be transferred without notice}%
{Shell \MakeLowercase{\textit{et al.}}: Bare Demo of IEEEtran.cls for IEEE Journals}
\maketitle

\begin{abstract}
In this paper, we investigate the  control of inverter-based resources (IBRs) for optimal voltage unbalance attenuation (OVUA). This problem is formulated as an optimization program under a tailored dq-frame, which minimizes the negative-sequence voltage at the point of common coupling (PCC) subject to the current, active power, synchronization stability, and feasibility constraints. The program is inherently nonconvex and intractable. So, to guarantee the optimality,  a rigorous optimality analysis is performed by leveraging analytical optimization. The analysis is divided into two cases: full mitigation  of VU and partial attenuation of VU. For the former case, we analytically solve the original program since the resultant VU is immediately deducible. For the latter one, directly solving the problem becomes very hard. Thus, we reformulate the program into an equivalent but more tractable form under certain conditions, by which the analytical optimum can be finally derived. It is found that the optimum trajectory  has three stages (O1--O3), depending on two critical boundary conditions (C1 and C2). Finally, we implement the optimum with a photovoltaic (PV)-storage system by developing an OVUA controller. The proposed approach is demonstrated by dynamic simulations under different VU conditions and is compared with several existing practices.

\end{abstract}

\section{Introduction}
\IEEEPARstart{V}oltage unbalance (VU) is one of the most common power quality issues, which can be induced by unbalanced load, asymmetrical transmission lines, transformer configuration, asymmetrical faults, etc.  VU is undesirable because it results in adverse effects on equipment as well as on power systems. The equipment, such as induction motors,
power electronic converters and adjustable speed drives, suffers from reduced efficiency  and decreased life under unbalanced conditions; power systems not only incur more losses and heating effects, but also become less stable \cite{VJA2001}. Thus, attenuating voltage unbalance is of great importance. 

Inverter-based resources (IBRs) are considered a powerful tool for voltage unbalance attenuation (VUA) enhancement in distribution grids or microgrids due to the flexibility of inverters. It has received extensive attention in recent years and a popular way to achieve this establishes on the well-known symmetrical component theory and the decoupled dual-sequence control of inverters. %Correspondingly, 

Rule-based control strategies for VUA have been well studied \cite{CPT2009,GJM2013,NT2015,CA2021,MYARI2008,SM2012,YJ2013,LTL2013,GF2015,MNR2017,AS2019,GMM2020,BAM2021}. In these methods, the negative-sequence voltage or voltage balance factor is regulated to a reference value (e.g., zero) or within a predefined range, resorting to droop (proportional) control \cite{CPT2009,GJM2013,NT2015,CA2021} or proportional-integral (PI) control \cite{MYARI2008,SM2012,YJ2013,LTL2013,GF2015,MNR2017,AS2019,GMM2020,BAM2021}. Moreover, coordination among multiple IBRs is addressed, so that the work stress on negative-sequence current support can be evenly shared \cite{CPT2009,GF2015,BAM2021}. 
These strategies are computationally cheap and easy to implement. However, most of them deal with mild VU caused by the unbalanced load during steady-state operation, whereas their performance under significant VU (e.g., owing to asymmetrical grid faults) is unclear. Another common limitation is suffering from suboptimality due to the heuristic nature; see \cite{YG2021} for the discussion regarding the suboptimality of droop rule. 
 
Driven by the large-scale deployment of IBRs, it is worthwhile to pursue better solutions for VUA in the sense that the control of converters will highly impact the behavior of a converter-dominated power system. In this regard, a few advanced control strategies have been developed in \cite{LK2007,NF2016,CA2018,SMA2019}, wherein the negative-sequence voltage at the point of common coupling (PCC) is minimized subject to inverter current limit. Although the optimal control solutions in \cite{LK2007,NF2016,CA2018,SMA2019} are conducted and implemented in different ways [based on the synchronous reference frame ($\rm dq$-frame) or stationary reference frame ($\alpha\beta$-frame)], they are consistent at bottom---\emph{IBRs behave as a controlled current source injecting the maximum allowable negative-sequence current, so that the resultant phase angle of negative-sequence PCC voltage is exactly in line with that of negative-sequence grid voltage; or equivalently, IBRs behave as a passive impedance of which the angle is same as that of the grid impedance.} Unlike \cite{CPT2009,GJM2013,NT2015,CA2021,MYARI2008,SM2012,YJ2013,LTL2013,GF2015,MNR2017,AS2019,GMM2020,BAM2021}, those optimal strategies can deal with significant VU; however, their optimality is \emph{conditional},  as will be elaborated in this paper,.  
\iffalse
\begin{figure}[t!]
    \centering
    \includegraphics[width=3.2in]{phasordiagram.pdf}
    \caption{Phasor illustration of O1 where $\dot{V}_g^-$ and $Z:=R+j\omega L$ are the negative-sequence grid voltage and impedance, $\dot{V}^-$ is the negative-sequence PCC voltage, and $\dot{I}^-$ is the negative-sequence IBR output current.}
    \label{phasor_optimalcontrol}
\end{figure}
\fi

\begin{figure*}
    \centering
    \includegraphics[width=5in]{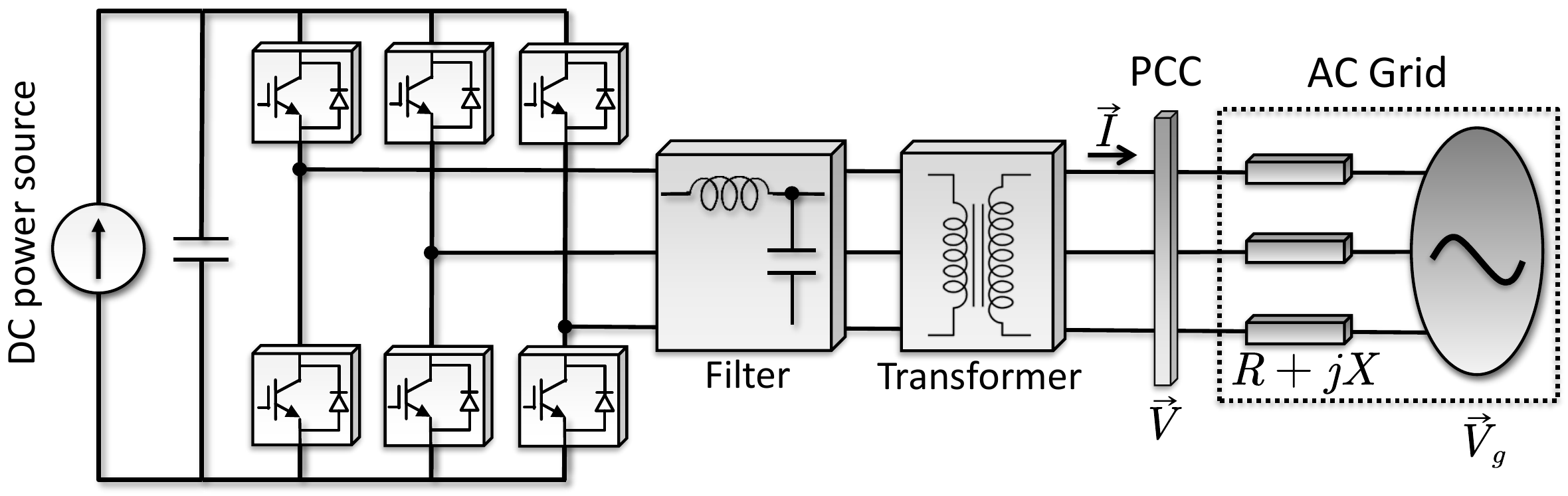}
    \caption{System configuration of a three-phase grid connected IBR system.}
    \label{systemconfig}
\end{figure*}

%  
%A few works propose compensating VU only relying on negative-sequence reactive current regulation \cite{NT2015,CA2021}; however, as we will argue, this cannot achieve a full mitigation except for a pure inductive network, even though the VU is very small. 

This paper aims to achieve optimal VUA (OVUA) using inverters under any given VU conditions (VU level and grid strength), resorting to optimization. So, the first question of significance is: how should it be formulated? Suggested by VU's definition in \cite{PP2001}, there is little doubt that reducing negative-sequence voltage is the most straightforward way for VUA in a three-phase three-wire system; we consider it as the goal in this work. As for constraints, it is observed that only the current limit is well considered in existing methods to avoid overloading of inverters, whereas the active power (at ac side) and synchronization stability constraints are not. Clearly, ignoring them may induce instability \cite{YG2021}.
First, without the active power limit, there is a risk of a power imbalance between the dc and ac sides; as a result, dc-link voltage will become unstable. The resultant active power delivery on the ac side may be infeasible for the dc side; so, the optimum in \cite{LK2007,NF2016,CA2018,SMA2019} may be physically infeasible. This is one reason why it is conditional. Such an issue is not given attention to except in \cite{CA2018}, where a suboptimal strategy is proposed for a special case---IBRs purely inject the maximum reactive current if active power absorption is not allowed. Second, as reported in \cite{GO2014}, \emph{loss of synchronism} (LoS, also termed as synchronization instability \cite{GH2018}, current angle instability \cite{WB2015}, etc.) of inverters may occur once the current phase angle is driven approaching the LoS boundary under severe (positive-sequence) voltage sags and correspondingly, the voltage support control is enhanced to avoid LoS \cite{GO2014,GH2018}. However, this instability is often left out in VUA studies. As elaborated later, LoS could also happen if the negative-sequence current is not adequately regulated. The conventional PI control, droop control, and suboptimal strategies \cite{CA2018,SMA2019} face with this threat. Third, there should be an additional consideration of the feasibility of power flow. Although in theory, IBRs can operate in four quadrants, some of the operating points (active and reactive currents) are, in fact, physically infeasible. Unlike the synchronization instability, such infeasiblity does not violate the LoS boundary, though physically, they both result in system oscillations. It is worth noting that the optimum given in \cite{LK2007,NF2016,CA2018,SMA2019} probably suffers from this infeasibility under mild VU and/or weak grid conditions. This is another reason why we argue that it is conditional. Thus, the feasibility of control solution should be guaranteed so that IBRs can operate with a stable equilibrium, which, to the best of our knowledge, has  not been addressed so far. Since the physical nature of LoS can  also be regarded as the infeasibility of power flow, it is expected to uniformly characterize such infeasibility and LoS.

Once the problem is formulated, the second question of interest is: can it be efficiently solved  with \emph{global} optimality guarantees? Unfortunately, the problem is not convex. This implies that numerical solvers usually do not offer optimality guarantees. It is worth noting that in \cite{LK2007,NF2016,CA2018,SMA2019}, the optimum is solved based on the Karush–Kuhn–Tucker (KKT) conditions, i.e., KKT point as optimum. Strictly speaking, KKT conditions only necessarily hold for regular local optima. Given that the additional nonlinear constraints should be considered besides the current limit, a comprehensive and rigorous optimality analysis is needed. Moreover, unlike steady-state optimization, there is a stringent demand for computational efficiency to suit the fast real-time control. To meet this requirement, an analytical solution would be preferred.

The contributions of this paper include three aspects: the thorough formulation, the rigorous optimality analysis as well as the robust implementation, which are detailed as follows: 
\begin{enumerate}
    \item We model the system based on the tailored dq-frame for negative-sequence system to reduce the complexity of formulation and analysis. To make the model tractable for optimization, it is simplified without loss of exactness by examining the feasibility of solutions. Then, an OVUA model that minimizes negative-sequence voltage and covers all the necessary physical constraints is proposed. 
    \item The optimality analysis of OVUA is performed using analytical optimization. Given the complexity, it is nontrivial to analytically solve OVUA with all constraints, simultaneously. Thus, we propose to  look for its exact relaxations  under different VU conditions. Leveraging the underlying geometry properties of the optimization program, we recognize the critical constraints that really matter  to construct an exact relaxation. This significantly reduces the program's complexity; thereby, all the candidates for optimum can be found analytically.
\item We implement the theoretical results with a  photovoltaic (PV)-storage system  in a combined open- and closed-loop fashion.
To guarantee robust synchronization of IBRs, the optimum is finally converted to the control solution under the dq-frame based on positive-sequence voltage.
\end{enumerate}

This paper improves the earlier studies  \cite{LK2007,NF2016,CA2018,SMA2019}  in the following ways. First, all the active power, synchronization stability, and feasibility constraints are correctly  addressed, so that IBRs can contribute to VUA while safely operating  themselves. Second, through the rigorous optimality analysis, the solution of OVUA under any given VU conditions is provided. Third, the analysis methods and results in this work can also achieve optimal positive-sequence voltage reduction during a high-voltage ride-through. Moreover, as a byproduct, this paper, together with our earlier work \cite{YG2021}, could provide some reference for IBRs planning, where the voltage support and/or VU compensation capabilities are specified.

The rest of this paper is organized as follows. Section II gives the problem statement of the OVUA problem, including the modeling and problem formulation. Section III provides the optimality analysis along with some geometric illustrations. Section IV proposes the implementation method with a PV-storage system. Simulation results are presented in Section V, followed by the conclusion. Most derivations and proofs are
collected in the Appendix.
 %Most derivations and proofs can be found in the arXiv version of this paper \cite{YG2021_2}.
\section{Problem Statement}
\subsection{System Model}
Fig. \ref{systemconfig} shows the system configuration of a three-phase grid-connected IBR system. The dc-side source is considered on a broader concept, which could be distributed generators (e.g., PV and wind), energy storage, controllable loads, or their combination. It is connected to the grid through a voltage-source inverter, an LC filter, and a step-up transformer; the grid side of the transformer is seen as the PCC bus. 

The typical control structure of a three-phase grid-connected inverter consists of an outer power/voltage control loop and an inner current control loop under the synchronous dq-reference frame. For VUA, the dual sequence control method that allows decoupled positive-and negative-sequence control is adopted; see \cite{YA2006} for more details. The  grid is modeled by the Thévenin equivalent, which emulates unbalanced grid conditions where $V_g$ and $R+jX$ denote the negative-sequence grid voltage and impedance, respectively.

Taking the phase sequence `B-A-C', the negative-sequence system can be treated as a \emph{virtual} positive-sequence system with Phase B being the leading phase; thereby, all the conventional  methods for positive-sequence analysis also hold for the virtual system, and this virtual system enjoys the same properties of the positive-sequence system regarding voltage magnitude, current magnitude and active power before and after the Park transformation. We advocate adopting such virtual system and the corresponding dq-frame ($\rm DQ^-$ for short) throughout the modeling and analysis because 1) it reduces the complexity of model expression and corresponding analysis, and 2) it helps reveal a particular dual relationship between the positive-sequence voltage support \cite{YG2021} and negative-sequence voltage attenuation. 
For simplicity of expression, any voltage, current, and power refer to the physical quantities associated with the fundamental frequency negative-sequence system throughout the paper, unless otherwise specified.

Accordingly, the fundamental frequency negative-sequence physics of inverters under $\rm DQ^-$ can be described as:
\begin{subequations}\label{PhModel}
\begin{align}
    I&=\sqrt{I_{d}^2+I_{q}^2} \\
 V&=\sqrt{V_{d}^2+V_{q}^2}=V_d\\ 
    P&=\frac{3}{2}\left(V_{ d}I_{d}+V_{ q}I_{ q}\right)=\frac{3}{2}VI_d
%    Q^-&=1.5\left(V_{\rm q}^-I_{\rm d}^--V_{\rm d}^-I_{\rm q}^-\right)=-1.5V_d^-I_q^-
\end{align}
\end{subequations}
where $V$, $I$, and $P$ are the voltage magnitude, current magnitude, and active power (average over one  fundamental frequency period) at the PCC; $V_{d}$ and $V_{q}$ are the d-axis and q-axis components of  PCC voltage; $I_{ d}$ and $I_{q}$ denote the active and reactive currents, respectively.

The current and active power limits of IBR are given as,
\begin{align}
\label{currentlimit} 0\leq I&\leq \overline{I}\\
\label{powerlimit} \underline{P}\leq P&\leq \overline{P}
\end{align}
where $\overline{I}>0$ denotes the current limit; $ \underline{P}\leq0$ and $\overline{P}>0$ are the lower and upper active power limits.

\subsection{Physics of Voltage Unbalance Attenuation }
 By Kirchhoff's voltage law, the negative-sequence network can be expressed as:
\begin{align}\label{powerflowlaw}
 \vv{V}-\vv{V_g}=\vv{I}\underbrace{(R+jX)}_{Ze^{j\phi}}
\end{align}
where $\vv{V_g}:=V_ge^{j0}, \vv{V}:=Ve^{j\theta}$ and $\vv{I}$ are the phasor representations of grid voltage, PCC voltage and IBR output current; $Z=\sqrt{R^2+X^2}$ and $\phi={\rm atan2}(X,R)$. Keep in mind that an implicit condition of (\ref{powerflowlaw}) is,
\begin{align}\label{Vnonneg}
    V\geq0.
\end{align}

To derive the relationship between the controllable variables $(I_d,I_q)$ and voltage solution $(V,\theta)$, multiply both sides of (\ref{powerflowlaw}) by $e^{-j\theta}$ and split the real and imaginary parts, which yields
\begin{subequations}\label{powerflowreim}
\begin{align}
V-RI_{d}+XI_{q}&=V_g\cos{\theta}\\
RI_{q}+XI_{d}&=V_g\sin{\theta}.
\end{align}
\end{subequations}

Based on (\ref{powerflowreim}), it follows that any  feasible power flow solution $(V,\theta)$ yields a specific control solution $(I_d,I_q)$ but the inverse is not valid. Therefore, some control solutions $(I_d,I_q)$ may not yield a feasible power flow solution $(V,\theta)$. 
\subsection{Optimal Voltage Unbalance Attenuation Model}
OVUA aims to find the optimal control solution of $(I_d,I_q)$ that minimizes negative-sequence voltage while satisfying the current, active power,  synchronization stability, and feasibility constraints. So, the problem can be formulated as
\begin{subequations}\label{P0}
\begin{align}
\hspace{-5mm}{\bf P0}:\hspace{2mm} \underset{}{\rm min}\hspace{3mm} &V\\
\notag{\rm over}\hspace{3mm}&I_{d},I_{q},V,\theta\\
{\rm s.t.}\hspace{3mm} &I_d^2+I_q^2\leq\overline{I}^2\\
&\underline{P}\leq\frac{3}{2}VI_d\leq\overline{P}\\
&V-RI_{d}+XI_{q}=V_g\cos{\theta}\\
&RI_{q}+XI_{d}=V_g\sin{\theta}\\
&V\geq0
\end{align}
\end{subequations}
where the first and second constraints are the current and active power limits; the last three constraints are the feasibility constraints that inherently cover the synchronization stability. In the analysis, we take $\theta\in[-180^\circ,180^\circ]$.

The following assumptions are made throughout this paper.
\begin{itemize}
    \item We consider a resistive-inductive grid, i.e., $R>0$ and $X>0$; the grid impedance is assumed to be available. However, in practical implementation, an online grid estimation impedance is preferred, which is out of the scope of this paper; some recent studies can be referred to \cite{TMG2020,MN2021}.
    \item The inner power losses of the IBR system are negligible in the optimality analysis. Still, a real lossy system is considered in the implementation and simulation, where a feedback mechanism is designed to deal with it.
    \item This paper focuses on the negative-sequence voltage reduction; the positive-sequence system performance is of no interest, and the positive-sequence current is considered zero without loss of generality. This implicitly requires that the negative-sequence control has a higher priority than positive-sequence control so that the current capacity of inverters can be fully exploited for VUA. The remaining available current and active power can be further utilized for positive-sequence control, e.g., positive-sequence voltage support \cite{YG2021}. If the positive-sequence control is given the higher priority, it may affect the net available current and active power for negative-sequence control, i.e., the current magnitude limit ($\overline{I}$) and (average) active power limits ($\overline{P}$ and $\underline{P}$) in P0. Once a specific positive-sequence control is determined, those parameters should be modified accordingly to avoid over-current and power imbalance\footnote{The net active power limits can be directly computed by the subtracting the components associated with the positive sequence from the total available active power. For the net current limit, a straightforward but conservative strategy is subtracting the positive-sequence current magnitude from the maximum allowable phase current magnitude \cite{NT2015}.}; however, note that the optimality analysis method in this paper is still valid after the modification. The optimal coordination between the positive and negative-sequence control is out of the scope of this paper, but it is an interesting topic worthy of investigation as future work.  
\end{itemize}

\section{Optimality Analysis}

In this section, the optimality analysis of P0 is conducted analytically, of which the benefits include:
\begin{itemize}
       \item it guarantees the global optimality;
    \item it reveals how the parameters ($V_g, R, X, \overline{I}, \underline{P}$ and $\overline{P}$) impact the  solution; and
    \item it can be used for designing a fast real-time controller.
\end{itemize}

\begin{figure*}
    \centering
    \includegraphics[width=2.7in]{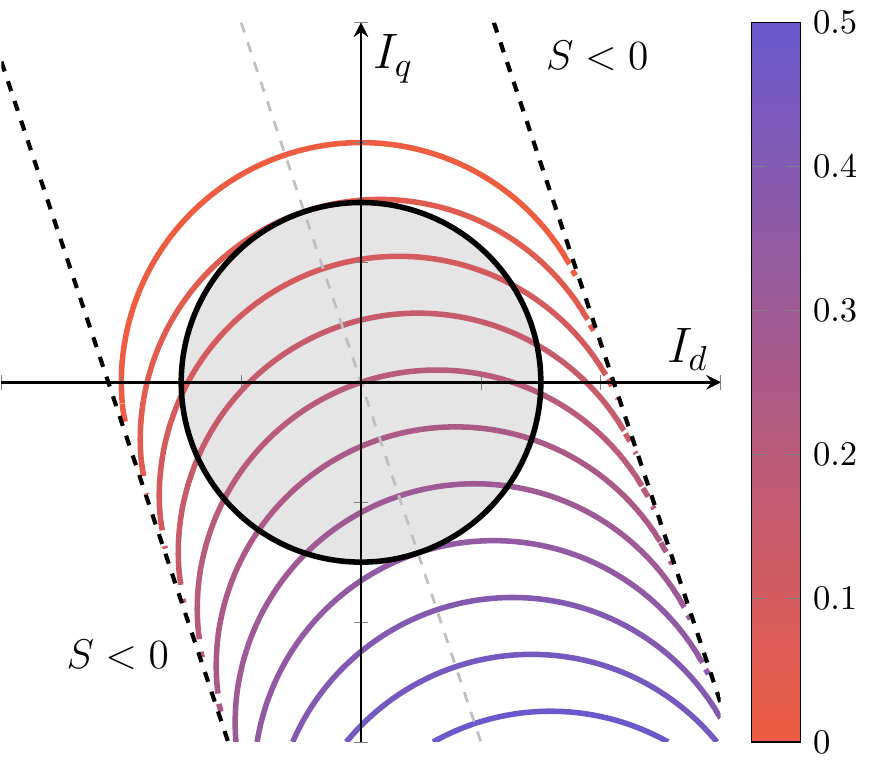}\hspace{22mm}
     \includegraphics[width=2.8in]{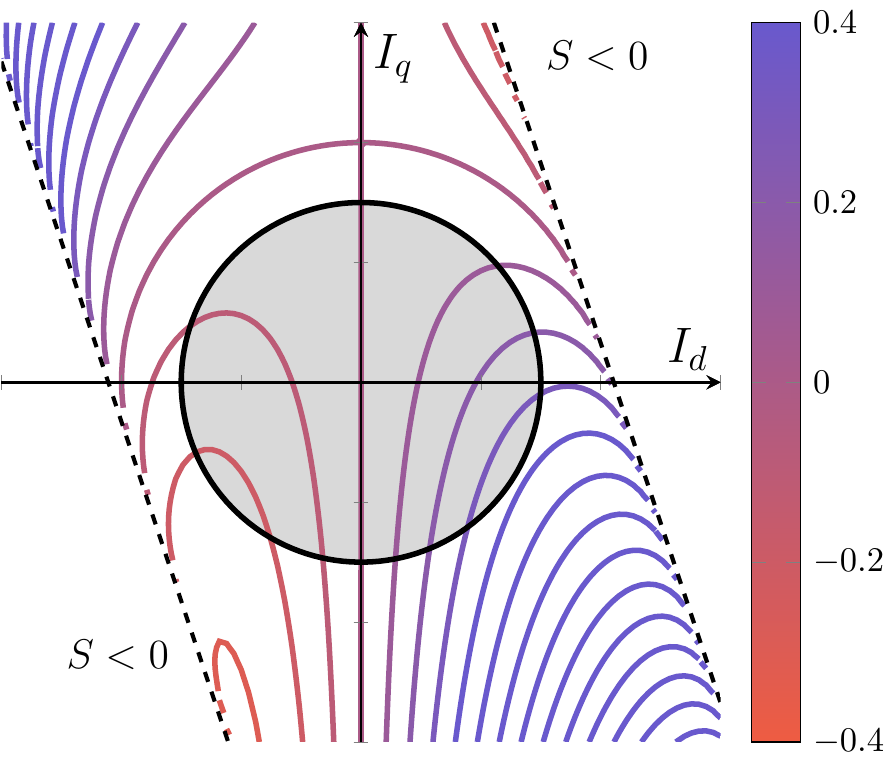}
    \caption{Contours of voltage (left) and active power (right) where $V_g=0.2$ pu, $Z=0.1$pu and $R/X=1/3$ in this example.}
    \label{Vcontour}
\end{figure*}

However, the complexity of P0 makes it very difficult to solve analytically. Hence, we will explore the conditions, under which the program can be  simplified and thus becomes tractable. Based on (\ref{P0}f), the lower bound of the cost function is $V=0$, which corresponds to the \emph{full mitigation} of VU. However, this may not be achievable under certain conditions. The following optimality analysis will start with full mitigation; then, we consider the cases with only \emph{partial} mitigation, where the problem can be reformulated into an equivalent form but tractable version. Note that all the solutions will be derived and expressed under $\rm DQ^-$.

\subsection{Full Mitigation of Voltage Unbalance}
The optimality of P0 under this case is elaborated as follows.

\emph{Theorem 1:} The full mitigation of VU can be achieved, i.e., ${V^{\star}}=0$ in P0, if and only if the following condition holds:
\begin{align}\label{C1}
    {\bf C1:}\hspace{5mm} \overline{I}\geq I_b:=\frac{V_g}{Z}
\end{align}
and ${\bf x}:=(I_d,I_q)^T$ is optimal if and only if,
\begin{align}\label{relaxO1}
    ||{\bf x}||_2=\sqrt{I_d^2+I_q^2}=\frac{V_g}{Z}.
\end{align}

Based on (\ref{powerflowlaw}), there is a unique current phasor solution: 
\begin{align}
    \vv{I}^\star=-\frac{\vv{V}_g}{Ze^{j\phi}}=I_be^{j(180^\circ-\phi)}
\end{align}
which yields $V^\star=0$. 
However, paradoxically, Theorem 1 implies that there should be an infinite number of solutions ${\bf x}^\star$ achieving the full mitigation of VU. This is because  only the current magnitude is restricted to the fixed value $I_b$, whereas the restriction on active current (by the active power constraint) is  relieved by $V^\star=0$. In this context, the reference dq-frame can be arbitrarily chosen in (\ref{powerflowlaw}), i.e., $\theta^\star$ is not unique; thereby, $\vv{I}^\star$  
corresponds to different  ${\bf x}^\star$ under different dq-frames.

To be in line with the implementation later in Section IV, we here take the grid voltage as the reference signal (i.e., $\theta^\star=0^\circ$) and the optimum is thus uniquely expressed as,
\begin{align}\label{O1}
{\bf O1}:\hspace{5mm}   {\bf x}^\star=\left(-\frac{R}{Z}I_b,\frac{X}{Z}I_b\right)^T.
\end{align}
\iffalse
\begin{figure}
    \centering
    \includegraphics[width=2.8in]{O1_trajectory.pdf}
    \caption{Geometrical illustration of the solution that achieves mitigation of VU where the red line corresponds to the solution satisfying (\ref{relaxO1}), the blue point is O1, the solid black denotes the current limit boundary and the dashed black line denotes synchronization stability boundary $|RI_d+XI_q|=V_g$.}
    \label{geo_O1}
\end{figure}

The geometric illustration of the optimum trajectory (\ref{relaxO1}) and  O1 is given as Fig. \ref{geo_O1}.
\fi

Theorem 1 implies that whether or not full mitigation can be achieved only depends on the  relationship between the current limit $\overline{I}$ and $I_b$. Considering $\overline{I}$ is fixed, full mitigation can be expected if the VU is very small and/or the grid is weak.

Moreover, the fact that the optimum in \cite{LK2007,NF2016,CA2018,SMA2019} may suffer from infeasibility under C1 is revealed below.

Such optimum under $\rm DQ^-$ corresponds to the solution
\begin{align*}
I_d=-\frac{R}{Z}\overline{I},\hspace{5mm}I_q=\frac{X}{Z}\overline{I}.\end{align*}
Substituting it into (\ref{powerflowreim}), one can obtain
\begin{align*}
V=-\overline{I}Z+V_g\cos{\theta}\leq-\overline{I}Z+V_g<0
\end{align*}
if C1 holds. This contradicts  (\ref{Vnonneg}). So, if it is implemented based on ${\rm DQ}^-$, the system will fail to reach an equilibrium. 

\subsection{Partial Attenuation of Voltage Unbalance}
The problem solving becomes much more complex in this case since $V^\star$ is no longer directly inferable. Moreover, the nonlinear constraints, especially for the existence of cosine and sine functions of $\theta$, make the program intractable. So, firstly, we  seek to simplify the program by eliminating $\theta$.

\emph{Lemma 1}: If C1 does not hold, then it follows that
\begin{enumerate}
    \item the optimum of P0 satisfies $-90^\circ\leq\theta^\star\leq90^\circ$, and
    \item P0 is an exact relaxation of the following program: 
\begin{subequations}\label{P0prime}
\begin{align}
\hspace{-5mm}{\bf P0}^\prime:\hspace{2mm} \underset{}{\rm min}\hspace{3mm} &V(I_d,I_q)\\
\notag{\rm over}\hspace{3mm}&I_{d},I_{q}\\
{\rm s.t.}\hspace{3mm} &I^2(I_d,I_q)\leq\overline{I}^2\\
&P(I_d,I_q)\geq\underline{P}\\
&P(I_d,I_q)\leq\overline{P}\\
&V(I_{d},I_{q})\geq0
\end{align}
\end{subequations}
where 
\begin{align}\label{VIdIq}
 \hspace{-5mm} V(I_d,I_q)=\sqrt{V_g^2-(RI_q+XI_d)^2}+RI_d-XI_q.
\end{align}  
\end{enumerate}
  
This implies that if C1 does not hold, P0 can be replaced by P0$^\prime$ without loss of optimality. The decision variables of P0$^\prime$ reduce to $I_d$ and $I_q$ without $\theta$, and thus, P0$^\prime$ becomes more tractable. We will equivalently solve P0$^\prime$ in the following analysis. And note that the optimal angle $\theta^\star\in[-90^\circ,90^\circ]$ can be uniquely recovered from the optimum of P0$^\prime$.
Besides, there is an underlying constraint in P0$^\prime$:
\begin{align}\label{stabilityregion}
    S(I_d,I_q):=|RI_q+XI_d|-V_g\leq0.
\end{align}

It is, in fact, the synchronization stability condition  for the negative-sequence system, which has a similar expression  as the counterpart for the positive-sequence system \cite{YG2021}. It guarantees the existence of a real value solution of $V$ mathematically and the equilibrium of the system physically. Any control solution that does not satisfy  (\ref{stabilityregion}) is physically infeasible and will induce LoS.

Before solving P0$^\prime$, it is formally clarified: the optimum of P0$^\prime$ exists because it minimizes a continuous function over a closed and bounded set. The proof regarding the existence of optimum is similar to the one for positive-sequence voltage maximization in \cite{YG2021} and omitted here for brevity.

Though P0$^\prime$ has a simpler form than P0, it is still nonconvex and is hard to be analytically solved with all the constraints simultaneously. Hence, we  further reduce the complexity by examining the geometry of P0$^\prime$. Fig. \ref{Vcontour} shows the   voltage and active power contours where the thick black line is the current limit boundary. Observing the contours, we can infer:
\begin{itemize}
\item The optimum appears in the second quadrant or on the $I_q$-axis.
    \item The upper bound of active power (\ref{P0prime}d) will not be binding at the optimum.
    \item The voltage constraint (\ref{P0prime}e) will not be binding at the optimum (because C1 does not hold).
\end{itemize}

Accordingly, we will explore the optimality of the relaxations of P0$^\prime$ only with the critical constraints (\ref{P0prime}b) and (\ref{P0prime}c) in the following analysis and find the sufficient conditions that guarantee the exactness. The following theorems state the main results.

\emph{Theorem 2:} Suppose C1 does not hold, the unique optimum of P0$^\prime$ is, 
\begin{align}
\label{O2}
 {\bf O2:}\hspace{3mm}   {\bf x}^\star=\left(-\frac{R}{Z}\overline{I},\frac{X}{Z}\overline{I}\right)^T
\end{align}
if the following condition holds:
\begin{align}
    {\bf C2:}\hspace{3mm}\underline{P}\leq P_b:=\frac{3}{2}\left(-\frac{R}{Z}V_g\overline{I}+\overline{I}^2R\right).
\end{align}

\emph{Theorem 3:} If neither of C1 and C2 hold, the unique optimum of P0$^\prime$ is,
\begin{align}
  {\bf O3:}\hspace{3mm}   {\bf x}^\star=\left(\overline{I}\cos\varphi^\star,\overline{I}\sin\varphi^\star\right)^T
\end{align}
which satisfies
\begin{align}
    P({\varphi}^\star)=\underline{P},\hspace{3mm}90^\circ\leq\varphi^\star\leq 180^\circ-\phi.
\end{align}

By Lemma 1, Theorems 2 and 3 also provide the optimum for P0. This implies that 1) the current limit is always binding at the optimum and 2) the lower bound of active power becomes the only key factor that affects the optimum trajectory if C1 does not hold. Note that O2 is, in fact, same as the obtained optimum in \cite{LK2007,NF2016,CA2018,SMA2019}. Theorem 2 reveals that it is indeed the global optimum only if C1 does not hold while C2 holds. In the last subsection, we have elaborated that if C1 holds, O2 is infeasible. Here, it is shown that if C2 does not hold and O2 is implemented, there will be a power imbalance between the dc and ac sides of inverters. This will lead to a significant boost of dc-link voltage and threaten the safety of IBR systems. O2 usually corresponds to the case that VU is sufficiently large and/or the grid is rigid while IBR can absorb considerable active power; on the contrary, O3 corresponds to the case that the allowable power absorption is somewhat limited. Besides, it is shown that O3 cannot be explicitly expressed, but as will be detailed later, it can be tracked in a closed-loop manner.
\begin{figure*}
    \centering
    \includegraphics[width=7in]{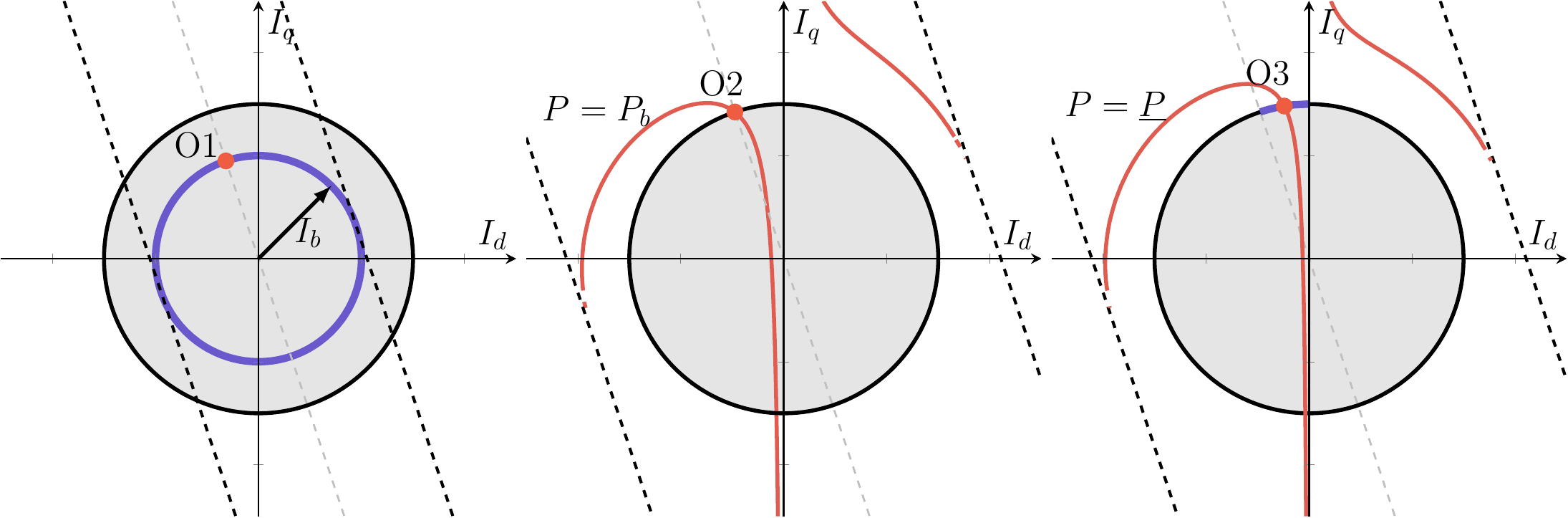}
    \caption{Geometric illustration of optimum where the black circle denotes the current limit boundary $I(I_d,I_q)=\overline{I}$, the orange curves denote the contours of active power at specific value, and dashed lines are the stability boundary $S(I_d,I_q)=0$.}
    \label{geometryOPT}
\end{figure*}
\subsection{Geometric Illustration of Optimum}
To better explain the underlying geometric properties, we here give the graphical illustration of the optimum trajectory  under different VU conditions. 

The left part of Fig. \ref{geometryOPT} corresponds to the case of VU full mitigation (i.e., C1 holds). The inner violet circle denotes the trajectory $I(I_d,I_q)=I_b$, and the orange point is O1. Geometrically, O1 is the second-quadrant intersection point between the trajectory (\ref{relaxO1}) and the following straight line:
\begin{align}
    \mathcal{L}1:\hspace{3mm}RI_q+XI_d=0.
\end{align}
In this case, the stability boundary has an intersection with the current limit boundary. This suggests that some solutions within the current limit may cause LoS and therefore become infeasible regardless of active power constraints. If the grid has a relatively high R/X ratio, the point $(0,\overline{I})$ could be out of the synchronization stability boundary, i.e., $S(0,\overline{I})>0$. This indicates that the strategy that injects the maximum allowable reactive current to reduce VU \cite{CA2018} will induce LoS.

The middle part corresponds to the case that O2 is the optimum ($\underline{P}\leq P_b$). O2 is the intersection between the current limit boundary and $\mathcal{L}1$ in the second quadrant. As $\underline{P}$ increases so that $\underline{P}>P_b$, the optimum O3 (right part of Fig. \ref{geometryOPT}) will move along the current limit boundary towards the point $(0,\overline{I})$. O3 is geometrically the unique intersection point between the curves $P(I_d,I_q)=\underline{P}$ and the current limit boundary in the second quadrant. For the cases that C1 does not hold, the synchronization stability boundary has no intersection with the current limit boundary, and therefore LoS will never happen.

\section{Implementation}
Theorems 1--3 have given the optimum of the OVUA problem. This section presents the implementation method of such optimal control solution, which mainly aims to deal with VU during short-term abnormal operations caused by asymmetrical faults.\footnote{The optimality analysis results in this work can also be  used for mitigating mild long-term VU during normal operation (e.g., caused by unbalanced loads). However, the trigger condition and the coordination between the positive and negative sequence control should be further investigated so that IBRs can contribute to VUA  while achieving certain primary operation goals, e.g., MPPT operation.} Without loss of generality, this paper considers a combination of PV and backup energy storage as the dc-side source, enabling active power absorption. The OVUA controller will be integrated into the IBR control system to generate current references for attenuating VU.
\subsection{Overview}
In normal operation, the IBR system operates in the maximum power point tracking mode with the unity power factor. When the OVUA control mode is triggered (any phase voltage $\leq0.85$pu), the following tasks will be carried out:
\begin{itemize}
    \item The dc voltage reference generated by the MPPT control is latched while the normal dc voltage controller is locked. Given that this work considers an energy storage system with small power rating, the PV panel will opt-out and energy storage opt-in during OVUA. The estimated maximum allowable power absorption ${\underline{P}}$ is initially set as per the power rating of the energy storage system.
    \item The grid model estimation is performed, which provides the grid parameters for the OVUA controller. The following simple strategy is used to estimate the grid voltage. During the first $m$ cycles ($m = 3\sim5$) after triggering OVUA, the positive and negative sequence dq current references are set as zero so that the grid voltage can be locally estimated. The information, including positive-sequence and negative-sequence grid voltage and the angle difference, can be estimated and used in the  control. 
    \item Based on the knowledge of grid conditions and IBR parameters, the negative-sequence current references under $\rm DQ^-$, i.e., $(I_d^\star, I_q^\star)$, are generated as per Theorems 1--3. 
\item Then, the generated current references $(I_d^\star, I_q^\star)$ are transformed to the counterpart $(I_d^{-\ast}, I_q^{-\ast})$ under the dq-frame based on the positive-sequence voltage ($\rm DQ^+$ for short), as the final input of the current control loop.
\item The optimality analysis is performed based on a lossless system. However, the active power losses can, in fact, enlarge the available power absorption capabilities of IBRs, which may further enhance the VUA performance. This indicates that the optimum can  possibly be promoted  from O3 to O2. So, if O3 is initially implemented, it will be rechecked if O3 can be promoted to O2 after obtaining the  active power measurements at the PCC. This is carried out at the $n$th cycle after implementing O3.
\end{itemize}

\subsection{OVUA Controller Design}
\begin{figure}
    \centering
    \includegraphics[width=2.5in]{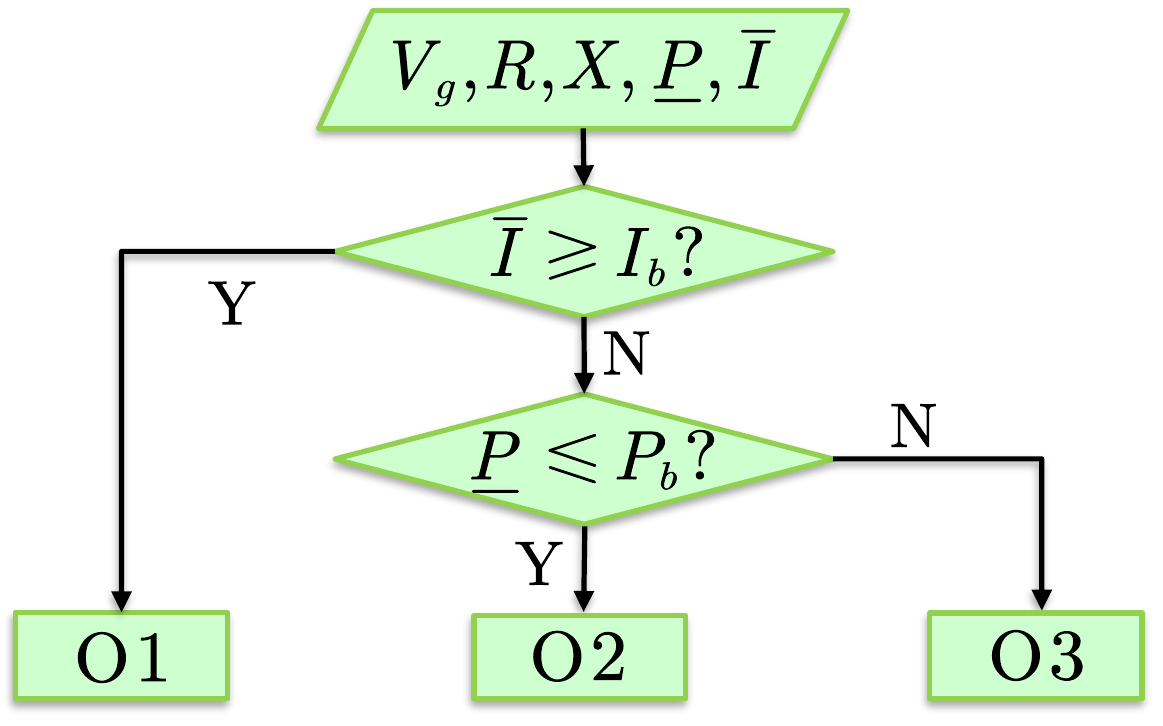}
    \caption{Logic loop of optimum selection.}
    \label{logicloop}
\end{figure}
\begin{figure*}
    \centering
    \includegraphics[width=6.5in]{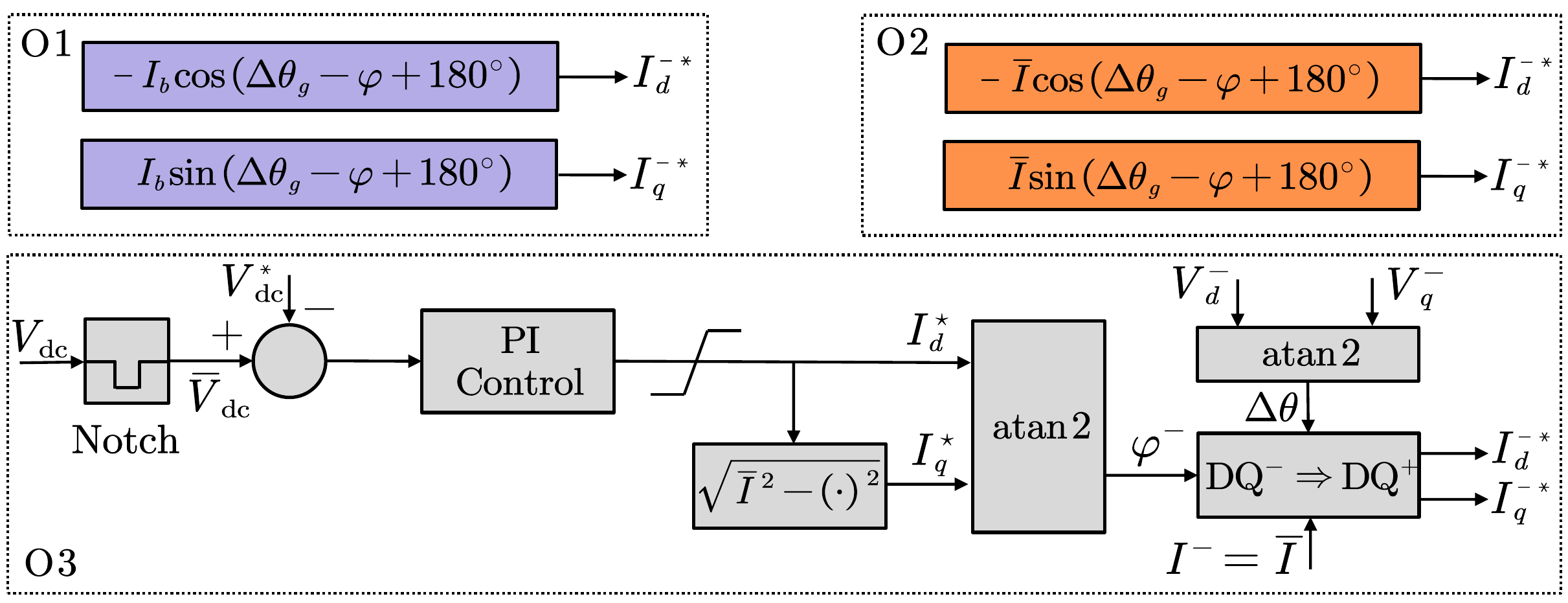}
    \caption{OVUA controller design.}
    \label{OVUAcontroller}
\end{figure*}
The logic loop that determines the active stage (O1/O2/O3) is illustrated in Fig. \ref{logicloop}. As mentioned above, this selection will be carried out twice to achieve the potential optimum promotion if O3 is initially activated. 

If the negative-sequence voltage is small (whether initially or after attenuation), the negative-sequence voltage angle cannot be computed robustly. This often happens when the VU can be fully mitigated. In that case, we use the positive-sequence voltage angle to create a robust dq-frame, in line with the typical dual sequence control. 

Let $(I_d^{\star},I_q^{\star})$ be the dq currents under $\rm DQ^-$ and $(I_d^-,I_q^-)$ be its counterpart under $\rm DQ^+$, respectively. Then, we have
\begin{align}\label{DQtrans}
   I_d^-&= -I^-\cos(\theta^--\theta^++\varphi^-)\\
     I_q^-&=     I^-\sin(\theta^--\theta^++\varphi^-)
\end{align}
where $\theta^+$ and $\theta^-$ are the phase angle of positive-sequence and negative-sequence voltage at the PCC, respectively; and
\begin{align}
    I^-&=\sqrt{(I_d^{\star})^2+(I_q^{\star})^2}\\
    \varphi^-&={\rm atan2}\left({I_q^{\star}},{I_d^{\star}}\right).
\end{align}

It can be observed that, besides $I^-$ and $\varphi^-$, the angle difference $\Delta\theta:=\theta^--\theta^+$ is required for the transformation. Given that the positive-sequence current is controlled to be zero, we have $\theta^+=\theta^+_g$.  However, as mentioned above, the negative-sequence voltage angle cannot be robustly estimated when VU is very small. So, for O1 and O2,  the dq-frame based on the negative-sequence  grid voltage ($\rm DQ^-_g$) will be leveraged as the intermediate to achieve the optimum because the angle difference $\Delta\theta_g:=\theta^-_g-\theta^+=\theta^-_g-\theta^+_g$ can be computed during the first $m$ cycles and memorized, where $\theta^+_g$ and  $\theta^-_g$ are the phase angle of positive-sequence and negative-sequence components of grid voltage. Given that the positive-sequence current is controlled to be zero, $\theta^+=\theta^+_g$.

Therefore, the OVUA controller
is designed as in Fig. \ref{OVUAcontroller}, which is detailed as follows:
\begin{itemize}
    \item If O1 is triggered, the dq current references are generated in an open-loop manner according to (\ref{O1}). So, we have
    \begin{align}
        I^{-}&=I_b\\
        \varphi^-&=180^\circ-\phi\\
 \Delta\theta&=\Delta\theta_g\label{O1angletrans}
    \end{align}
where (\ref{O1angletrans}) holds because O1 in (\ref{O1}) is precisely  conducted based on $\rm DQ^-_g$ ($\theta^\star=0$). $\Delta\theta$ is computed at the end of the $m$th cycle.
\item As for O2, by (\ref{O2}) and (\ref{powerflowreim}b), one has $\theta^\star=0$. This indicates that $\rm DQ^-$ is aligned with $\rm DQ^-_g$ when the optimum is achieved. Accordingly, one can obtain
    \begin{align}
        I^{-}&=\overline{I}\\
        \varphi^-&=180^\circ-\phi\\
 \Delta\theta&=\Delta\theta_g.
    \end{align}
\item If O3 is triggered, it implies that there will be a considerable VU even after the attenuation. In this case, the negative-sequence voltage angle can be well estimated. To be consistent with O1 and O2 and thereby reduce the complexity of the control system, we first compute the current reference under $\rm DQ^-$ and then transform it to the counterpart under $\rm DQ^+$.
Since O3 does not have an explicit form, it cannot be directly implemented in an open-loop fashion as O1 and O2. However, as per Theorem 3, the inverter should inject the maximum current and the dc source should absorb power as much as possible. So, a PI controller for dc-link voltage is exploited to regulate the active current while the reactive current reference is computed correspondingly. Hence, the dq current references under $\rm DQ^-$ are given by, 
\begin{align}
  {I}_d^\star&=K_p\left(\bar V_{dc}-V_{dc}^\ast\right)+K_i\int \left(\bar V_{dc}-V_{dc}^\ast\right) dt\\
    {I}_q^\star&=\sqrt{\overline{I}^2-({I}_d^\star)^2}
\end{align}
where $\bar{V}_{dc}$ and $V_{dc}^\ast$ denote the filtered dc voltage and dc voltage reference; $K_p$ and $K_i$ are the proportional and integral coefficients. Then, we have
    \begin{align}
        I^{-}&=\overline{I}\\
        \varphi^-&={\rm atan2}\left({{I}_q^\star},{I}_d^\star\right)\\
\Delta\theta&={\rm atan2}\left(-{V_q^-},{V_d^-}\right)
    \end{align}
where $V_d^-$ and $V_q^-$ are the instantaneous dq-axes components of negative-sequence PCC voltage under $\rm DQ^+$.
Different than O1 and O2, here $\Delta\theta$ and $\varphi^-$ are estimated in real time. 
\end{itemize}

Remember  that the transformation from $\rm DQ^-$ to $\rm DQ^+$ is unique; and the voltage magnitude, current magnitude,  and instantaneous/average active power do not change through the transformation. This implies that the physical constraints satisfied under $\rm DQ^-$ are still satisfied under $\rm DQ^+$. Therefore, such transformation does not affect the OVUA performance. More  details about the transformation from $\rm DQ^-$ to $\rm DQ^+$ DQ+ are
offered in the Appendix. %are offered in in the arXiv version of this paper. 

\section{Validation}
\begin{table}[t]
\small
\centering
\caption{System Parameters}\label{mainparameters}
\renewcommand\arraystretch{0.95}
\begin{tabular}{ m{2in} m{1in}}
        \hline\hline
Description&Value\\
         \hline
Power rating of PV& 250 kW (1 pu) \\
Nominal ac voltage& 250V/25kV (1 pu)\\
Nominal dc voltage&480V\\
Nominal frequency& 60 Hz\\
Pre- and post-fault SCRs& 20/10\\
R/X ratio&2\\
Maximum current limit& 1.5 pu\\
Number of delayed cycles $m,n$& 3, 20\\
%PV module& SunPower SPR-415E-WHT-D \\
 \hline\hline
        \end{tabular}
\end{table}
This section tests the proposed OVUA with a grid-connected PV-storage system (as shown in Fig. \ref{systemconfig}) via dynamic simulations in MATLAB/Simulink R2021a environment. More specifications of the test system are provided in Table \ref{mainparameters}. The energy storage is modeled as a controlled power source  in the dynamic simulation. Four cases are considered for testing our proposal under different operation conditions (different voltage sags and different power ratings of backup energy storage). Before the unbalanced voltage sag occurs, the PV system operates with the rated power. We compare OVUA with the conventional droop control \cite{NT2015}, PI control \cite{LTL2013,GMM2020}, and the (sub)optimal voltage unbalance attenuation control strategy in \cite{CA2018} (VUA for short). For the PI control and droop control, the pure negative-sequence reactive current support is provided. The control solutions of those strategies are computed based on $\rm DQ^-$ and  finally implemented under $\rm DQ^+$.
\subsection{Full Mitigation of Voltage Unbalance: O1}
\begin{figure}
    \centering
        \includegraphics[width=3.5in]{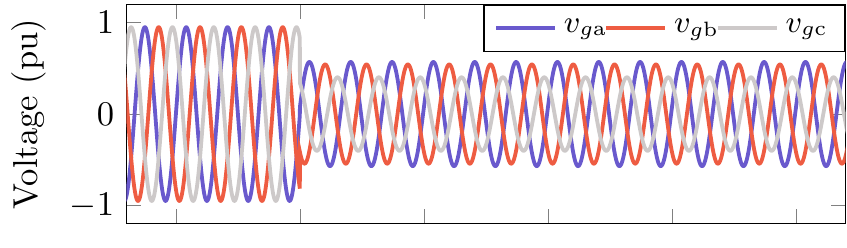}
        \includegraphics[width=3.5in]{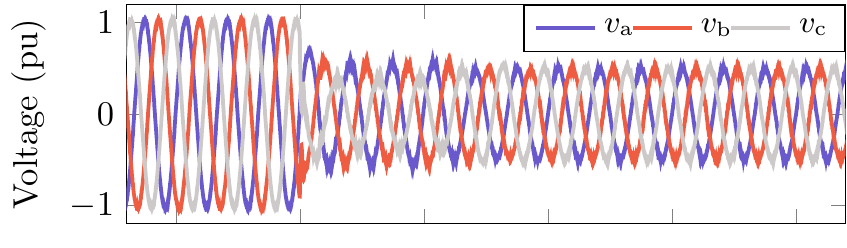}
         \includegraphics[width=3.5in]{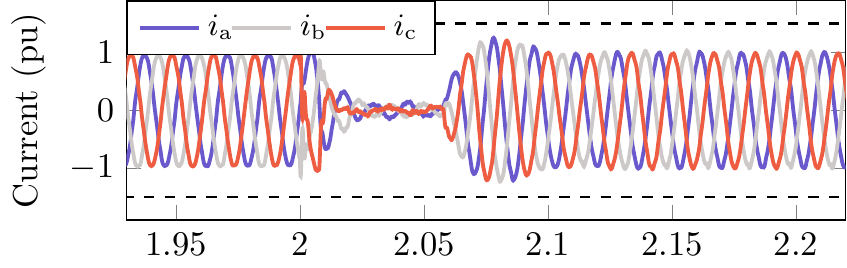}
    \caption{Simulation results with OVUA under Cases A1 and A2 (from top to bottom: three-phase grid voltage, PCC voltage, and output current).}
    \label{figs_caseA}
\end{figure}

In the following Cases A1 and A2, an unbalanced voltage sag that occurs at $t=2$s is imitated, where the positive-sequence voltage is $0.5\angle0^\circ$pu and the negative-sequence grid voltage is set as $0.1\angle50^\circ$pu. Since $I_b=1.0 {\rm pu}<1.5{\rm pu}=\overline{I}$, the OVUA strategy should operate with O1. Since O1 is not affected by the available power absorption, the control solutions of OVUA are the same in Cases A1 and A2. The three-phase grid voltage, PCC voltage, and current are shown in Fig. \ref{figs_caseA}.

{\bf Case A1.} In this case, there is no energy storage available at the dc side; therefore, active power absorption is not allowed, i.e., $\underline{P}=0$. This enables a fair comparison among PI control, droop control, VUA (wherein the suboptimal strategy is triggered), and OVUA. It can be observed that PI control and VUA both result in large oscillation (LoS instability) because PI control and VUA both drive the system towards the operation point $I_d=0,I_q=\overline{I}$ (under $\rm DQ^-$), which is not a feasible solution of (\ref{powerflowreim}) [and $S(I_d,I_q)>0$]. Thus, the system fails to reach an equilibrium. The system can converge to a stable equilibrium with droop control by selecting a conservative droop gain,  but the VU mitigation is very limited. If the droop gain is too large, the system will also become unstable. In comparison, the proposed OVUA not only guarantees stability but also fully mitigates the VU. 
\begin{figure}
    \centering
        \includegraphics[width=3.5in]{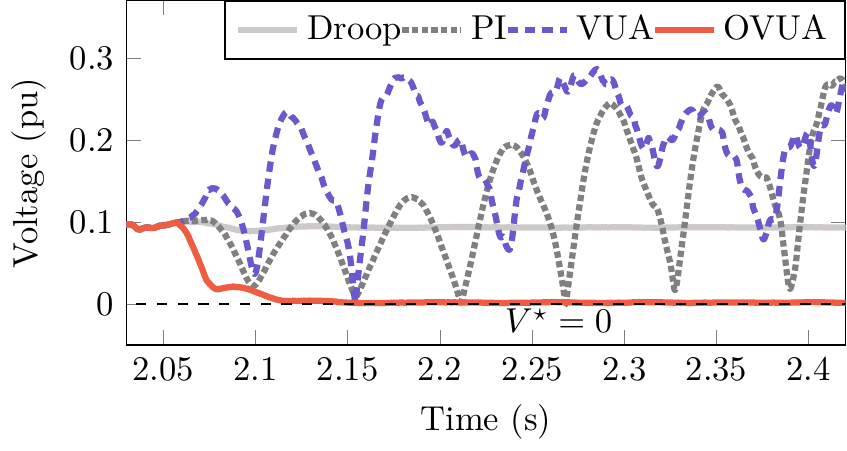}
      \caption{Negative-sequence voltage magnitude at the PCC under Case A1.}
    \label{Vneg_caseA1}
\end{figure}

\begin{figure}
    \centering
        \includegraphics[width=3.5in]{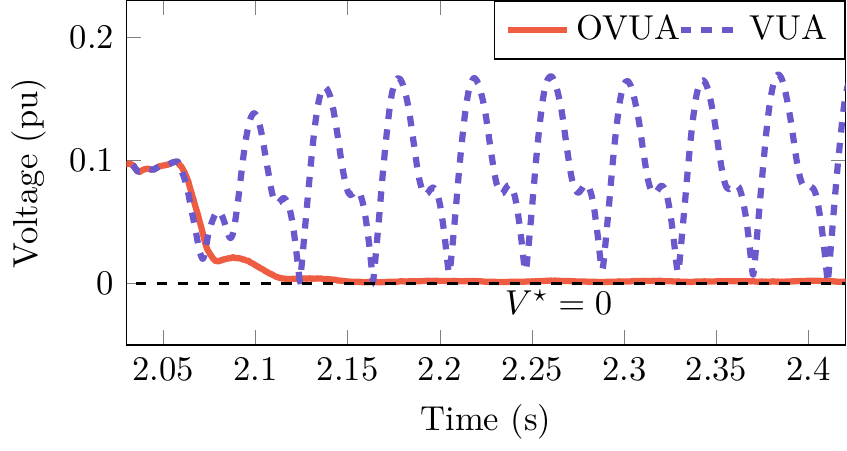}
      \caption{Negative-sequence voltage magnitude at the PCC under Case A2.}
    \label{Vneg_caseA2}
\end{figure}

{\bf Case A2.} In this case, suppose there is a considerable active power absorption capability of IBR --- the power rating of energy storage is set as $\rm 0.3pu$. In this context, VUA operates with the optimal strategy, i.e., $I_d=-(R/Z)\overline{I}$ and $I_q=(X/Z)\overline{I}$ (under $\rm DQ^-$). The simulation result is shown in Fig. \ref{Vneg_caseA2}. It shows that with VUA, the system still fails to attain an equilibrium. As proven in Section III-A, this is because such a control solution is physically infeasible. Unlike in A1, this solution satisfies $S(I_d,I_q)\leq0$; this instability is slightly different from the LoS but the rationales  behind them are same---both of the control solutions fail to yield a feasible solution for  Kirchhoff's voltage law (\ref{powerflowlaw}). 
\subsection{Partial Attenuation of Voltage Unbalance: O2}
\begin{figure}[t]
    \centering
        \includegraphics[width=3.5in]{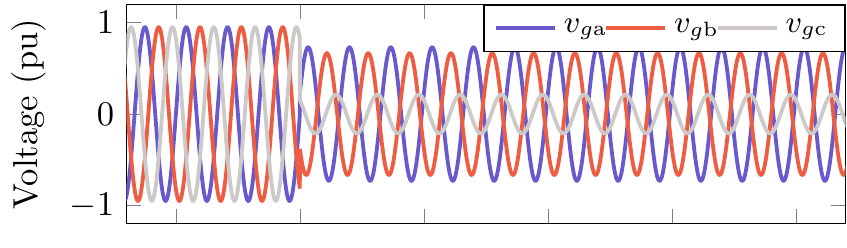}
        \includegraphics[width=3.5in]{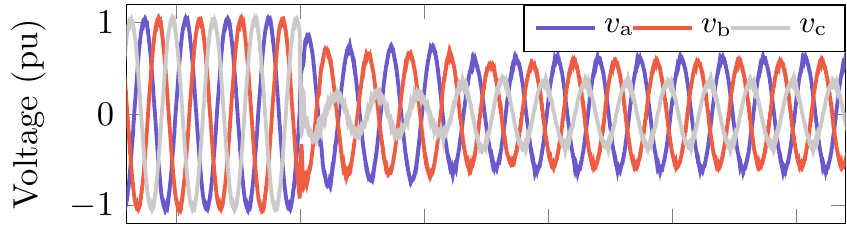}
         \includegraphics[width=3.5in]{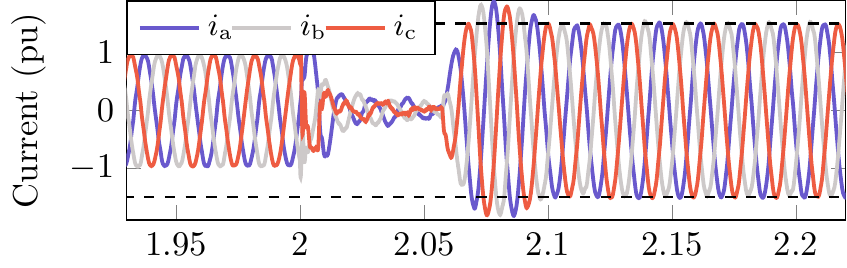}
    \caption{Simulation results with OVUA under Case B (from top to bottom: three-phase grid voltage, PCC voltage, and output current).}
    \label{figs_caseB}
\end{figure}
\begin{figure}
    \centering
        \includegraphics[width=3.5in]{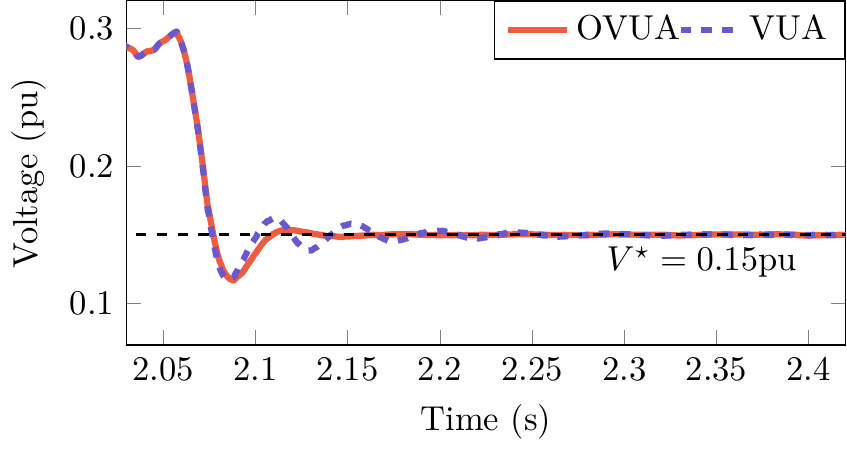}
      \caption{Negative-sequence voltage magnitude at the PCC under Case B.}
    \label{Vneg_caseB}
\end{figure}

In Case B, an unbalanced voltage sag that occurs at $t=2$s is imitated, where the positive-sequence voltage is $0.5\angle0^\circ$pu and the negative-sequence grid voltage is set as $0.3\angle50^\circ$pu. The power rating of energy storage is $\rm 0.3pu$.
Since $I_b=3 {\rm pu}>1.5{\rm pu}=\overline{I}$ and $P_b=-0.2{\rm pu}>-0.3{\rm pu}=\underline{P}$,  OVUA should operate with O2. The three-phase grid voltage, PCC voltage, and output current are shown in Fig. \ref{figs_caseB}. Fig. \ref{Vneg_caseB} gives the negative-sequence voltage at the PCC with the VUA and OVUA strategies. Both of them achieve the optimal VU attenuation where the negative-sequence PCC voltage is reduced from $0.3 {\rm pu}$ to $0.15{\rm pu}$ because they generate the same control solution in this case. The only difference is that OVUA reaches the equilibrium faster than VUA since OVUA exploits the \emph{apriori} open-loop angle information $\Delta\theta_g$  for O2.

\subsection{Partial Attenuation of Voltage Unbalance: O3}
\begin{figure}[t]
    \centering
        \includegraphics[width=3.5in]{vgabc_caseB.pdf}
        \includegraphics[width=3.5in]{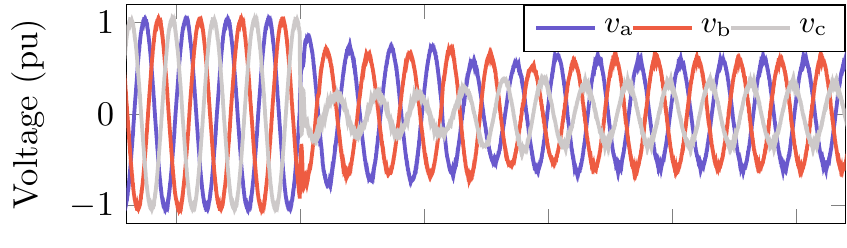}
         \includegraphics[width=3.5in]{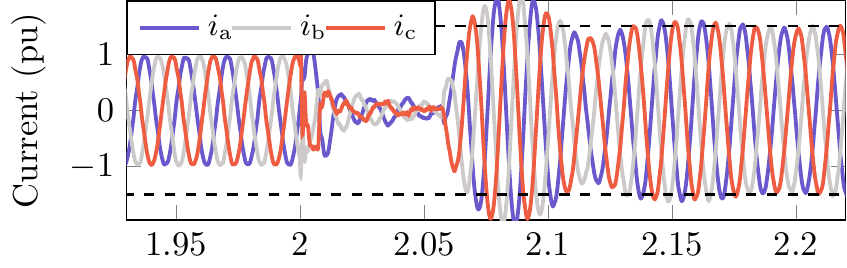}
    \caption{Simulation results with OVUA under Case C (from top to bottom: three-phase grid voltage, PCC voltage, and output current).}
    \label{figs_caseC}
\end{figure}
\begin{figure}
    \centering
        \includegraphics[width=3.5in]{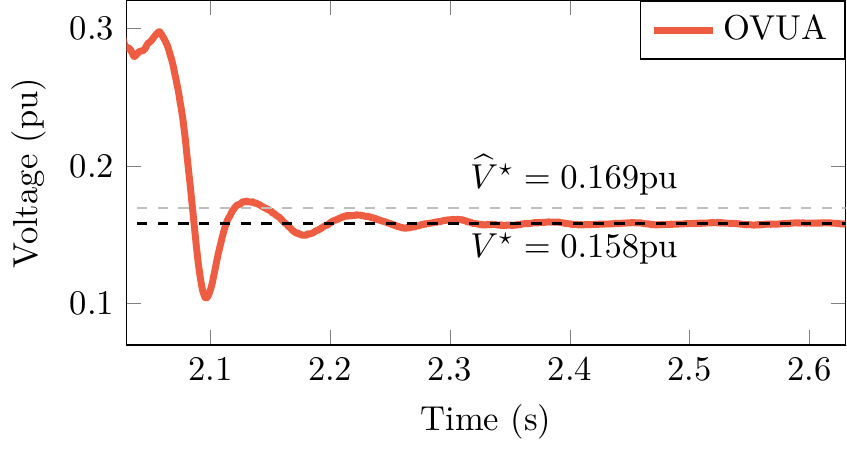}
      \caption{Negative-sequence voltage  magnitude at the PCC under Case C.}
    \label{Vneg_caseC}
\end{figure}
\begin{figure}[t]
    \centering
        \includegraphics[width=3.5in]{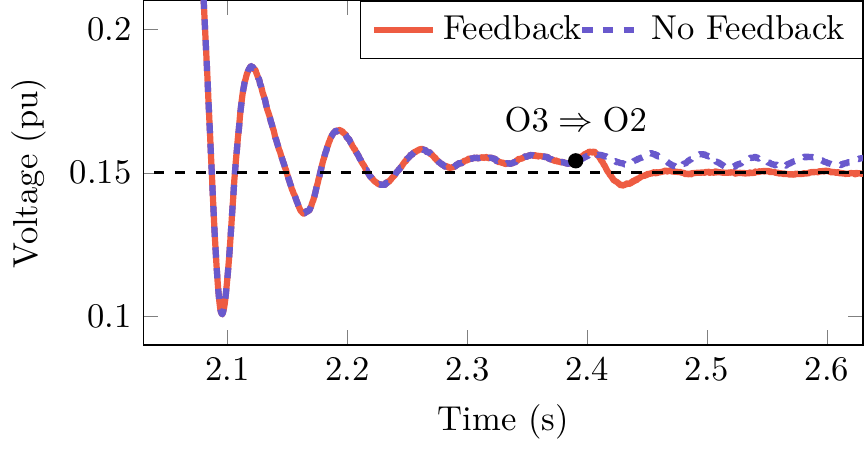}
      \caption{Negative-sequence voltage magnitude at the PCC under Case D.}
    \label{Vneg_caseD}
\end{figure}
\begin{figure}[h!]
    \centering
        \includegraphics[width=3.5in]{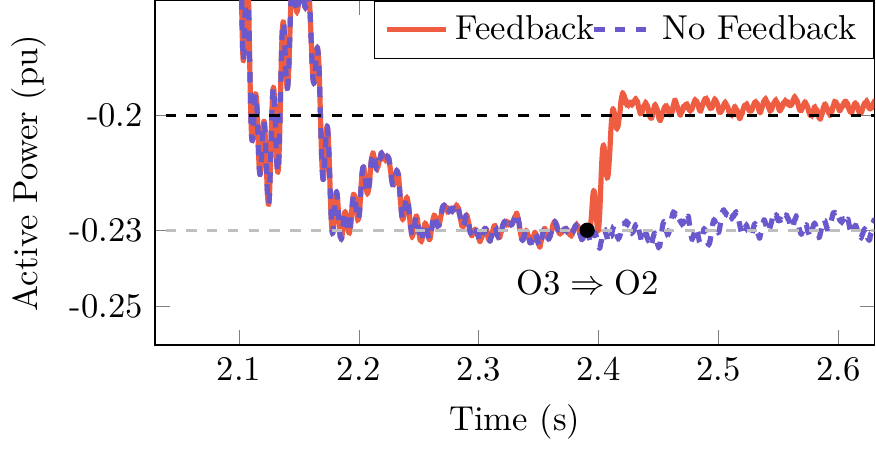}
      \caption{Filtered active power measurement at the PCC under Case D.}
    \label{Pac_caseD}
\end{figure}
In Case C, the same unbalanced voltage sag that occurs at $t=2$s is imitated as in Case B. Differently, the power rating of energy storage is  $\rm 0.1pu$. Given that $I_b=3 {\rm pu}>1.5{\rm pu}=\overline{I}$ and $P_b=-0.2{\rm pu}<-0.1{\rm pu}=\underline{P}$, OVUA should operate with O3. Fig. \ref{figs_caseC} illustrates the three-phase grid voltage, PCC voltage and current. Fig. \ref{Vneg_caseC} shows the negative-sequence voltage at the PCC. It is observed that the real resultant VU ($V^\star=0.158{\rm pu}$) is slightly smaller than the expected VU ($\widehat V^\star=0.169{\rm pu}$) computed based on $\underline{P}=-0.1{\rm pu}$ due to inner power losses of the system. Since the VUA strategy does not specify the control solution under this condition, no comparison can be provided here. 

\subsection{Promotion From O3 to O2}
In Case D, the same VU condition is considered as in Case C. The only difference is the power rating of energy storage is set as $\rm 0.19pu$. Given that $I_b=3 {\rm pu}>1.5{\rm pu}=\overline{I}$ and $P_b=-0.2{\rm pu}<-0.19{\rm pu}=\underline{P}$, O3 should be triggered at the beginning. Figs. \ref{Vneg_caseD} and \ref{Pac_caseD} show the negative-sequence voltage and filtered active power at the PCC. OVUA initially operates with O3; then, after $20$ cycles, it switches to O2 due to the feedback mechanism. The active power measurement (filtered by Notch) is around $-0.23\rm pu$, which is actually lower than $P_b=-0.2\rm pu$ due to the existence of power losses. Compared with the non-feedback strategy, the system with a feedback mechanism quickly reaches the new better optimal operation point (the same one as in Case B) after the transition.

\section{Conclusion}
This paper investigates the optimal voltage unbalance attenuation provided by IBRs. We model the OVUA problem under a tailored dq frame, where negative-sequence voltage is minimized subject to the physical constraints in terms of current magnitude, active power, synchronization stability, and feasibility. To solve this nonlinear nonconvex program with global optimality guarantees, we perform a rigorous optimality analysis where the analytical optimum is found. Then, we implement it with a PV-storage system by developing an OVUA controller, of which  the effectiveness is well validated. It shows better control performance compared with the existing PI control, droop control as well as the  optimal strategies. Some facts are for the first time revealed in this paper:
\begin{enumerate}
    \item Under given VU conditions, there always exists a unique optimum for OVUA, but its expression is not uniform. The trajectory of optimal solutions  has three stages (O1--O3), and it depends on two key factors: inverter current limit and maximum allowable active power absorption. The  boundary conditions are  given by C1 and C2.
    \item The negative-sequence control may also cause LoS instability when the VU is small and/or the external grid is weak.  The existing methods (PI control, droop control, and the VUA strategy) may suffer from  instability. 
    \item The active power losses inside the IBR system can further improve the VUA performance since power losses enlarge the active power absorption capability.
\end{enumerate}

The current implementation method requires a grid model estimation with an acceptable accuracy, which might not be straightforward; so, we will explore some advanced implementation strategies  based on the optimum trajectory, which do not rely on the model estimation. Besides, the optimal coordination between positive and negative-sequence control and the coordination among multiple IBRs at the network level will be addressed in the future work.

\appendices
\section{Proof of Theorem 1}
\emph{Proof of Sufficiency:}  The solutions satisfying (\ref{relaxO1}) can be written as,
\begin{align}\label{OptP0_fullmitigation}
I_d(\theta):=\frac{V_g}{Z}\cos(\phi+\theta),\,\, I_q(\theta)=-\frac{V_g}{Z}\sin(\phi+\theta)
\end{align}
where $ \theta\in[-180^\circ,180^\circ]$.
Then, we have,
\begin{itemize}
    \item $\sqrt{I_d^2(\theta)+I_q^2(\theta)}={V_g}/{Z}<\overline{I}$ given that C1 holds---(\ref{P0}b) is satisfied;
    \item $1.5VI_d(\theta)=0$, which satisfies (\ref{P0}d); and
    \item  any pair of $(I_d(\theta),I_q(\theta))$ along with $V=0$ and $\theta$ should satisfy (\ref{P0}d)--(\ref{P0}f).
\end{itemize}

Therefore, a full mitigation of voltage unbalance is achievable and any $\bf x(\theta)$ as in (\ref{OptP0_fullmitigation}) is an optimum of P0.\qed

\emph{Proof of Necessity:}  Given that $V^\star=0$ and the optimum must be feasible, by (\ref{P0}d)--(\ref{P0}e), we can obtain
\begin{align*}
(I_d^\star)^2+(I_q^\star)^2=\left(\frac{V_g}{Z}\right)^2\leq\overline{I}^2.
\end{align*}

This completes the proof.\qed
\section{Proof of Lemma 1}
\emph{Proof of Part 1):} Suppose there is an optimum of P0 satisfying
$$
|\theta^\star|>90^\circ,\,\, \theta^\star\in[-180^\circ,180^\circ].
$$
Then, based on (\ref{powerflowlaw}) and $V^\star>0$, we have
\begin{align*}
    I^\star=\frac{1}{Z}\sqrt{(V^\star)^2+V_g^2-2V^\star V_g\cos\theta^\star}>\frac{V_g}{Z}.
\end{align*}
This indicates that C1 holds, which contradicts the assumption. 

Therefore, we have $|\theta^\star|\leq 90^\circ$.
\qed

\emph{Proof of Part 2):} Based on $|\theta^\star|\leq 90^\circ$ and (\ref{powerflowreim}d)--(\ref{powerflowreim}e), it follows that
\begin{align}
 \notag  V^\star=\sqrt{V_g^2-(RI_q^\star+XI_d^\star)^2}+RI_d^\star-XI_q^\star.
\end{align}
This indicates that the optimum of P0 should satisfy (\ref{VIdIq}). Thus, P0 is an exact relaxation of P0$^\prime$.\qed

\section{Proof of Theorem 2}
The proof of Theorem 2 follows from Claims 1 and 2.

\emph{Claim 1:} Consider the following relaxation of P0$^\prime$ that only considers the current limit:
\begin{align*}
\hspace{-5mm}{\bf P1}:\hspace{2mm} \underset{}{\rm min}\hspace{3mm} &V(I_d,I_q)\\
{\rm s.t.}\hspace{3mm} &I^2(I_d,I_q)\leq\overline{I}^2.
\end{align*}
The optimum of P1 is
\begin{align*}
    {\bf x}^{\star}_{1}=\left(-\dfrac{R}{Z}\overline{I},\dfrac{X}{Z}\overline{I}\right)^T.
\end{align*}

\emph{Proof:} P1 is  non-convex given that $V$ is not convex. To find the global optimum, we will traverse  all the KKT points, irregular points, as well as nondifferentiable points. 

{\bf KKT Solution.} 
The Lagrangian of P1 is given as,
\begin{align*}
    L(I_d,I_q,\lambda):=V(I_d,I_q)+    \lambda\left(I_{d}^2+I_{q}^2-\overline{I}^2\right).
\end{align*}

Then, the first-order KKT conditions are:
\begin{subequations}\label{KKTONC1}
\begin{align}
 R-\dfrac{X(RI_{q}+XI_{d})}{\sqrt{V_g^2-(RI_{q}+XI_{\rm d})^2}}+2\lambda I_{d}&=0\\[0.5mm]
   -X-\dfrac{R(RI_{q}+XI_{d})}{\sqrt{V_g^2-(RI_{q}+XI_{d})^2}}+2\lambda I_{q}&=0\\
    I_{d}^2+I_{q}^2-\overline{I}^2&\leq0\\
    \lambda&\geq0\\ 
    \lambda\left(I_{d}^2+I_{q}^2-\overline{I}^2\right)&=0
\end{align}
\end{subequations}

 To derive the closed-form solution of (\ref{KKTONC1}), i.e., KKT points, we discuss two potential cases based on (\ref{KKTONC1}b): 1) $\lambda=0$ and 2) $\lambda>0$, where the second case implies that the constraint (\ref{KKTONC1}a) must be binding.

\begin{itemize}
    \item \emph{Case 1:} If $\lambda=0$, then from (\ref{KKTONC1}d) and (\ref{KKTONC1}e), we have 
\begin{align*}
    R^2=\dfrac{RX(RI_{q}+XI_{d})}{\sqrt{V_g^2-(RI_{q}+XI_{d})^2}}=-X^2
\end{align*}
which holds if and only if $R=X=0$. This contradicts our earlier assumption.
 \item \emph{Case 2:} If $\lambda>0$, $I_{d}^2+I_{q}^2-\overline{I}^2=0.$  
Multiplying (\ref{KKTONC1}d) and (\ref{KKTONC1}e) with $I_{q}$ and $I_{d}$, respectively, it follows that
\begin{align*}
    RI_{q}+XI_{d}=\dfrac{(XI_{q}-RI_{d})(RI_{q}+XI_{d})}{\sqrt{V_g^2-(RI_{q}+XI_{d})^2}}.
\end{align*}

So, Case 2 is further split into two subcases:
\begin{itemize}
\item \emph{Case 2a:}
\begin{align*}
    RI_{q}+XI_{d}\neq0 \Longrightarrow V_g\equiv\overline{I}Z.
\end{align*}
\item \emph{Case 2b:}
\begin{align*}
    RI_{ q}+XI_{d}=0  \Longrightarrow \left\{\hspace{-2mm}\begin{array}{l}
I_{d}=-({R}/{Z})\overline{I}\\
I_{q}=(X/{Z})\overline{I}\\
\lambda={Z}/({2\overline{I}}).
\end{array}\right.
\end{align*}
\end{itemize}
\end{itemize}

So, there is a unique KKT point $${\bf x}^{KKT}=\left(-\dfrac{R}{Z}\overline{I},\dfrac{X}{Z}\overline{I}\right)^T.$$

{\bf Irregular Solution.}
It is found that the unique solution of $\nabla I({\bf x})=0$, i.e., $(0,0)^T$,  does not make the inequality constraint $I^2({\bf x})-\overline{I}^2\leq0$ active. This means that it is still regular. Therefore, there is no irregular solution for P1. 

{\bf Nondifferentiable Solution.}
Given that C1 does not hold, there is no nondifferentiable point  on the constraint set of P1.

In summary, the unique KKT solution is  the optimum of P1, i.e., ${\bf x}^{\star}_{1}={\bf x}^{KKT}$. This completes the proof of Claim 1.
\qed

\emph{Claim 2:} P1 is an exact relaxation of P0$^\prime$ if C1 does not hold and C2 holds.

\emph{Proof:} 
{\bf Lower Bound of Active Power.} The solution of P1 yields
\begin{align*}
    P^\star&=P({\bf x}_1^\star)=\dfrac{3}{2}\left(-\dfrac{R}{Z}V_g\overline{I}+\overline{I}^2R\right).
\end{align*}

Since C2 holds, the constraint (\ref{P0prime}c) is satisfied.

{\bf Upper Bound of Active Power.} Since C1 does not hold, it follows that
\begin{align*}
P^\star=\dfrac{3}{2}R\overline{I}\left(-\dfrac{V_g}{Z}+\overline{I}\right)<0<\overline{P}.
\end{align*}
So, the constraint (\ref{P0prime}d) is satisfied.

{\bf Feasibility.} Since C1 does not hold, it follows that 
\begin{align*}
    V^\star=V_g-Z\overline{I}>0.
\end{align*}

The constraint (\ref{P0prime}e) is also satisfied. Therefore, P1 is an exact relaxation of P0$^\prime$ and the solution of P0 is ${\bf x}^{KKT}$. This completes the proof of Claim 2. \qed

\section{Proof of Theorem 3}
We first define the following auxiliary programs that will be used in the proof.
\begin{align*}
\hspace{-5mm}{\bf P2}:\hspace{2mm} \underset{}{\rm min}\hspace{3mm} &V(I_d,I_q)\\
{\rm s.t.}\hspace{3mm} &I^2(I_d,I_q)\leq\overline{I}^2\\
&P(I_d,I_q)\geq\underline{P}.
\end{align*}
\begin{align*}
\hspace{-5mm}{\bf P3}:\hspace{2mm} \underset{}{\rm min}\hspace{3mm} &V(I_d,I_q)\\
{\rm s.t.}\hspace{3mm} &P(I_d,I_q)=\rho.
\end{align*}
\begin{align*}
\hspace{-5mm}{\bf P4}:\hspace{2mm} \underset{}{\rm min}\hspace{3mm} &V(I_d,I_q)\\
{\rm s.t.}\hspace{3mm} &I^2(I_d,I_q)\leq\overline{I}^2\\
&P(I_d,I_q)=\rho.
\end{align*}

Theorem 3 follows from Claims 3 and 4.

\emph{Claim 3:} If neither of C1 and C2 hold, the the unique optimum of P2 is 
\begin{align}
 \notag {\bf x}^\star_2=\left(\overline{I}\cos\varphi^\star,\overline{I}\sin\varphi^\star\right)^T
\end{align}
which satisfies
\begin{align}
 \notag   P({\varphi}^\star)=\underline{P},\hspace{3mm}90^\circ\leq\varphi^\star\leq 180^\circ-\phi.
\end{align}

 Lemmas 2--4 are used in the proof of Claim 3.

\emph{Lemma 2:} Suppose $\rho>\underline{\rho}:=-3V_g^2/(8R)$, the optimum of P3 is describe as follows. 

If $({X^2}/{Z^2})\underline{\rho}\leq \rho\leq0$, 
\begin{align}
\nonumber    {\bf x}^\star_{3}=&\left(\frac{1}{2Z}\left(\dfrac{ X}{Z}V_g-\sqrt{\left(\dfrac{ X}{Z}V_g\right)^2+\frac{8}{3}R\rho}\right) ,\right.\\
  \notag     &\left.-\frac{X}{2RZ}\left(\dfrac{ X}{Z}V_g-\sqrt{\left(\dfrac{ X}{Z}V_g\right)^2+\frac{8}{3}R\rho}\right)+\frac{V_g}{R}\right)^T.
\end{align}

If $\underline{\rho}<\rho<({X^2}/{Z^2})\underline{\rho}$,
\begin{align}
  \notag  {\bf x}^\star_{3}=&\left(-\frac{1}{2Z}\left(V_g+\sqrt{V_g^2+{\frac{8}{3}R\rho}}\right)\right.,\\
   \notag   &\left.-\frac{X}{2RZ}\left(V_g-\sqrt{V_g^2+{\frac{8}{3}R\rho}}\right)\right)^T.
\end{align}

\emph{Proof:}
Similarly, since P3 is nonconvex, we will traverse  all the KKT points, irregular points, and nondifferentiable points to find the solution of P3. 

{\bf KKT Solution.} The Lagrangian of P3 is
\begin{align*}
  L(I_d,I_q,\lambda)=V(I_d,I_q)+\mu\left(P(I_d,I_q)-\rho\right)
\end{align*}
where $\mu$ is the Lagrange multiplier. 

The KKT conditions of P3 are then given as,
\begin{subequations}\label{KKT-ONC3}
\begin{align}
    \frac{3}{2}VI_{d}-\rho&=0\\
    \frac{\partial V}{\partial I_{d}}+\frac{3}{2}\mu\left(I_{d}\frac{\partial V}{\partial I_{d}}+V\right)&=0\\
    \frac{\partial V}{\partial I_{q}}+\frac{3}{2}\mu I_{d}\frac{\partial V}{\partial I_{q}}&=0.
\end{align}
\end{subequations}

So, based on (\ref{KKT-ONC3}b), there are two cases to be discussed.

\emph{Case 1:} Suppose ${\partial V}/{\partial I_{\rm q}}\neq0$, it is known that $I_{d}\neq0$ by checking (\ref{KKT-ONC3}c). And thus, $\mu={-2}/{(3I_{d})}$. Then, combining with (\ref{KKT-ONC3}b), we have $V=0$. This contradicts (\ref{KKT-ONC3}a), given that $\rho<0$. Therefore, there is no KKT solution in this case.

\emph{Case 2:} Suppose ${\partial V}/{\partial I_{q}}=0$, we have 
\begin{align}
    X+\frac{R(RI_{q}+XI_{d})}{\sqrt{V_g^2-(RI_{q}+XI_{d})^2}}=0\Rightarrow RI_{q}+XI_{d}+\frac{XV_g}{Z}=0\label{partialIq=0}.
\end{align}

Thus, the KKT points are the intersection points of (\ref{KKT-ONC3}a) and (\ref{partialIq=0}), which yields
\begin{align*}
    Z^2I_{d}^2+ZV_gI_{d}-\frac{2}{3}R\rho=0.
\end{align*} 

By solving it, we obtain two KKT solutions,
\begin{align*}
   {\bf x}^{KKT,1}=&\left(\frac{1}{2Z}\left(-V_g+\sqrt{V_g^2+\frac{8}{3}R\rho}\right)\right.,\\
   &\left.-\frac{X}{2RZ}\left(V_g+\sqrt{V_g^2+\frac{8}{3}R\rho}\right)\right)^T\\
     {\bf x}^{KKT,2}=&\left(-\frac{1}{2Z}\left(V_g+\sqrt{V_g^2+{\frac{8}{3}R\rho}}\right)\right.,\\
   &\left.-\frac{X}{2RZ}\left(V_g-\sqrt{V_g^2+{\frac{8}{3}R\rho}}\right)\right)^T
\end{align*}

By (\ref{KKT-ONC3}a), ${\bf x}^{KKT,2}$ is no worse than ${\bf x}^{KKT,1}$.

{\bf Irregular Solution.} By checking the linear independence constraint qualification, it is found that the minimizer cannot be \emph{irregular}. The reason is elaborated is as follows. The solutions of $\nabla P({\bf x})=0$ are
\begin{align*}
{\bf x}=\left(0,\frac{V_g}{Z}\right)^T, {\bf x}=\left(-\frac{V_g}{2Z},-\frac{XV_g}{2RZ}\right)^T
\end{align*}
The first solution is infeasible since C1 does not hold. The second solution is feasible only when $\rho=\underline{\rho}$, which contradicts the assumption $\rho>\underline{\rho}$.

{\bf Non-Differentiable Point.} We additionally check the non-differentiable points over the constraint set. There are four candidates whose $I_{d}$ are:
\begin{align*}
 I_{d1}&=\frac{1}{2Z}\left(-\dfrac{X}{Z}V_g+\sqrt{\left(\dfrac{ X}{Z}V_g\right)^2+\frac{8}{3}R\rho}\right)\\
 I_{d2}&=\frac{1}{2Z}\left(-\dfrac{X}{Z}V_g-\sqrt{\left(\dfrac{ X}{Z}V_g\right)^2+\frac{8}{3}R\rho}\right)\\
I_{d3}&=\frac{1}{2Z}\left(\dfrac{X}{Z}V_g+\sqrt{\left(\dfrac{ X}{Z}V_g\right)^2+\frac{8}{3}R\rho}\right)\\
I_{d4}&=\frac{1}{2Z}\left(\dfrac{X}{Z}V_g-\sqrt{\left(\dfrac{ X}{Z}V_g\right)^2+\frac{8}{3}R\rho}\right).
\end{align*}
Obviously, if 
\begin{align}
\notag    \left(\dfrac{X}{Z}V_g\right)^2+\frac{8}{3}R\rho\geq0\Rightarrow\rho\geq-\frac{X^2}{Z^2}\underline{\rho}>\underline{\rho}
\end{align}
the solutions must exist. Clearly, $I_{d3}\geq I_{d4}>0$. Thus, both of them yield a negative cost function value, definitely better than the KKT points that yield a
positive cost function value. Therefore, $I_{d4}$ corresponds to the optimum. 

In summary, if $({X^2}/{Z^2})\underline{\rho}\leq \rho\leq0$, the optimum is
\begin{align*}
\notag   {\bf x}^\star_{3}=&\left(\frac{1}{2Z} \left(\dfrac{ X}{Z}V_g-\sqrt{\left(\dfrac{ X}{Z}V_g\right)^2+\frac{8}{3}R\rho}\right),\right.\\
 \notag    &\left.-\frac{X}{2RZ}\left(\dfrac{ X}{Z}V_g-\sqrt{\left(\dfrac{ X}{Z}V_g\right)^2+\frac{8}{3}R\rho}\right)+\frac{V_g}{R}\right)^T.
\end{align*}
If $\underline{\rho}<\rho<({X^2}/{Z^2})\underline{\rho}$, ${\bf x}^\star_{3}={\bf x}^{KKT,2}$.
This completes the proof of Lemma 2. \qed

\emph{Lemma 3:} Suppose $\rho\geq P_b$, P3 has a unique optimum. 

\emph{Proof:} Define $$\delta(\overline{I}):=P_b-\underline{\rho}=\frac{3}{2}\left(-\dfrac{R}{Z}V_g\overline{I}+\overline{I}^2R\right)+\frac{3V_g^2}{8R}$$
which can be regarded as a quadratic function of $\overline{I}$. 

Let $\delta(\overline{I})=0$, we have
\begin{align}
 \notag \overline{I}^2R-\dfrac{R}{Z}V_g\overline{I}+\frac{V_g^2}{4R}=0.
\end{align}
It has solution(s) if and only if,
\begin{align}
  \notag  \frac{R^2}{R^2+X^2}V_g^2-V_g^2\geq0\Rightarrow \frac{X^2}{R^2+X^2}\leq0
\end{align}
which contradicts the assumption that $R>0,X>0$. In other words,
$0>P_b>\underline{\rho}$ always holds. Therefore, based on Lemma 2, if ${\rho}\geq P_b$, P3 has a unique optimum which completes the proof of Lemma 3. \qed

\emph{Lemma 4:} If neither of C1 and C2 hold, 
the solution space
\begin{align}
 \notag    \left\{{\bf x}\big|I({\bf x})=\overline{I},\, P({\bf x})=\underline{P}\right\}
\end{align}
is non-empty and there is at least one solution ${\bf x}={\bf x}(\varphi):=\big(\overline{I}\cos{\varphi},\overline{I}\sin{\varphi}\big)^T$ satisfying
\begin{align}
 \notag  P\left({\bf x}(\varphi)\right)=\underline{P}, \hspace{2mm}90^\circ<\varphi<{\rm atan2}\left({X},-{R}\right).
\end{align}

\emph{Proof:}
Since the solution satisfies $I({\bf x})=0$, we represent ${\bf x}$ in the polar coordinates, i.e., ${\bf x}(\varphi)=\left(\overline{I}\cos{\varphi},\overline{I}\sin{\varphi}\right)^T$.

If C1 does not hold, all the points within current limit circle is physically attainable and
$$
{\bf x}(90^\circ)=\left(0,\overline{I}\right)^T, {\bf x}(180^\circ-\phi)=\left(-\frac{R}{Z}\overline{I},\frac{X}{Z}\overline{I}\right)^T
$$
are the two feasible points, where $P({\bf x}(90^\circ))=0$ and $P({\bf x}(180^\circ-\phi))=P_b$, respectively. 

Then, since $P$ is continuous on $\varphi$, by \emph{intermediate value theorem}, there exists at least one solution $90^\circ<\varphi<180^\circ-\phi$, such that, $P({\bf x}(\varphi))=\underline{P}$ if C2 does not hold. This completes the proof of Lemma 4.\qed

\emph{Proof of Claim 3:} The core of this proof is to show that the solution of P2  makes both the constraints binding, that is, $P({\bf x}_2^\star)=\underline{P}$ and $I({\bf x}_2^\star)=\overline{I}$.

We first show that  $P({\bf x}_2^\star)=\underline{P}$ holds.

Suppose $P({\bf x}_2^\star)>\underline{P}$, then we can  relax the active power constraint in P2 without affecting the solution, and therefore P2 reduces to P1. This indicates that ${\bf x}_2^\star={\bf x}_1^\star$ and $$P({\bf x}_2^\star)=P({\bf x}_1^\star)=P_b$$ which contradicts the assumption (C2 does not hold). Therefore,  we have
\begin{align}
\label{bindinggp}
  P({\bf x}_2^\star)=\underline{P}.
\end{align}

Now, we show the necessity of $I({\bf x}_2^\star)=\overline{I}$.

Clearly, based on Lemma 2, the solution of P3 is a function of $\rho$, denoted as ${\bf x}_3^\star(\rho)$. Similarly, the solution of P4 is also related to $\rho$, denoted as ${\bf x}_4^\star(\rho)$.

Consider $\rho=P_b$. Since P1 is a relaxation of P4, ${\bf x}_1^\star$ is also feasible for P4. Thus, it follows that  ${\bf x}_{4}^\star(P_b)={\bf x}_1^\star$. P3 is also a relaxation of P4. Suppose $I({\bf x}_{3}^\star(P_b))<\overline{I}$. Then, since ${\bf x}_{3}^\star(P_b)$ is also feasible for P4, ${\bf x}_{3}^\star(P_b)$ is the optimum of P4. That implies that ${\bf x}_{3}^\star(P_b)={\bf x}_{4}^\star(P_b)={\bf x}_1^\star$. Thus, $I({\bf x}_{3}^\star(P_b))=I({\bf x}_{1}^\star)=\overline{I}$. This contradicts the assumption $I({\bf x}_{3}^\star(P_b))<\overline{I}$. Therefore, we have $I({\bf x}_{3}^\star(P_b))\geq\overline{I}$.

Consider $P_b<\rho\leq0$. Based on Lemma 3, P3 has a unique optimum, of which the corresponding $I_d$ is denoted as  $I_d({\bf x}_3^\star)$.

1) If $({X^2}/{Z^2})\underline{\rho}\leq\rho\leq0$, from Lemma 2, we have $V({\bf x}_3^\star)<0<V({\bf x}_1^\star)$. Given that P1 has a unique minimizer ${\bf x}_1^\star$, ${\bf x}_3^\star(\rho)$ must be infeasible for P1. So, $I({\bf x}_3^\star(\rho))>\overline{I}$.

2) If $P_b<\rho<({X^2}/{Z^2})\underline{\rho}$, based on Lemma 2, we have
$I_d({\bf x}_3^\star(\rho))<I_d({\bf x}_3^\star(P_b))<0$. So, it follows that $V({\bf x}_3^\star)<V({\bf x}_3^\star(P_b))$. Given that P1 has a unique optimum ${\bf x}_1^\star$, ${\bf x}_3^\star(\rho)$ must be infeasible for P1. So, $I({\bf x}_3^\star(\rho))>\overline{I}$.

Therefore, $I({\bf x}_3^\star(\rho))>\overline{I}$ always holds for any $P_b<\rho\leq0$ if  C2 does not hold. So, we have 
\begin{align}\label{bindinggc}
    I({\bf x}_3^\star(\underline{P}))>\overline{I}
\end{align}
if C2 does not hold.

Return to P2. Suppose the current constraint is not binding in P2, i.e., $I({\bf x}_2^\star)<\overline{I}$. Since it has been proven that $P({\bf x}^\star_2)=\underline{P}$, P2 can thus reduce to P3 with $\rho=\underline{P}$ via relaxing the inactive current constraint. This implies that ${\bf x}_2^\star={\bf x}_3^\star(\underline{P})$, and thus we have $I({\bf x}_2^\star)=I({\bf x}_3^\star(\underline{P}))>\overline{I}$ by (\ref{bindinggc}). However, this contradicts our earlier assumption. Therefore, the current limit constraint is always binding in P2, i.e., $I({\bf x}_2^\star)=\overline{I}$. 

Since $I({\bf x}_2^\star)=\overline{I}$ and $P({\bf x}_2^\star)=\underline{P}$, ${\bf x}_2^\star$ belongs to the solution space specified in Lemma 4. In Lemma 4, it is not guaranteed that the solution satisfying $P({\bf x}(\varphi))=\underline{P}$ is unique, though it is probably true. In case of multiple solutions, let  $\varphi^\star$ be the one with the smallest $I_{d}$ among all the solutions. Below, we will show that ${\bf x}_2^\star={\bf x}(\varphi^\star)$.

By (\ref{VIdIq}), it follows that,
\begin{align}\label{Vcircle2}
    \left(I_{d}-\frac{VR}{Z^2}\right)^2+\left(I_{q}+\frac{VX}{Z^2}\right)^2=\left(\frac{V_g}{Z}\right)^2.
\end{align}

Regard $(I_{d},I_{q})$ as a point in the rectangular coordinate, the trajectory of (\ref{Vcircle2}) is  a circle, of which the center is located at $(VR/Z, -VX/Z)$ exactly on the line $RI_{q}+XI_{d}=0$ and the radius is equal to $V_g/Z$. The  physically feasible trajectory of ($I_{d},I_{q}$) under  given $V$ is in fact the half circle that is over the line $RI_{d}-XI_{q}-V=0$. When this circle has overlap with the current limit boundary circle  (i.e., two circles are intersecting or tangent), the voltage is attainable within the current limit. As $V$ increases, the circle center will move along the line $RI_{q}+XI_{d}=0$ away from the point $(0,0)$. The intersection points are always symmetrical about the line $RI_{q}+XI_{d}=0$.

Assume there is another better solution than $\varphi^\star$ on the current limit boundary, it must be located within $(\varphi^\star,2(180^\circ-\phi)-\varphi^\star)$ because only this range has intersection points with a voltage circle that achieves lower $V$.

Since ${\bf x}(\varphi^\star)$ corresponds to the smallest $I_{\rm d}$ while satisfying $P({\bf x}(\varphi^\star))=\underline{P}$, there is  no solution $\varphi\in(\varphi^\star, 180^\circ-\phi]$ satisfying $P({\bf x}(\varphi))=\underline{P}$ because $\cos{\varphi}<\cos{\varphi^\star},\forall \varphi\in(\varphi^\star,180^\circ-\phi)$. Besides, there is also no solution $\varphi\in(\varphi^\star,180^\circ-\phi]$ satisfying $P({\bf x}(\varphi))>\underline{P}$. 
This is because if there is a solution $\varphi\in(\varphi^\star,180^\circ-\phi]$ satisfying $P({\bf x}(\varphi))>\underline{P}$, it can achieve lower voltage without making the minimum power constraint binding. This contradicts  (\ref{bindinggp}).

Now, we show that there is no better solution $\varphi\in(180^\circ-\phi,2(180^\circ-\phi)-\varphi^\star)$ as well. Assume there is a solution $\varphi\in(180^\circ-\phi,2(180^\circ-\phi)-\varphi^\star)$ satisfying $P({\bf x}(\varphi))=\underline{P}$. Its symmetrical point $\tilde\varphi$ regarding the line $RI_d+XI_q=0$ satisfies   $\tilde\varphi\in(\varphi^\star,180^\circ-\phi)$. Clearly, it follows that
\begin{align}\label{cosbound2}
 \notag  V({\bf x}(\varphi))=V({\bf x}(\tilde\varphi)),\, \cos{\varphi}<\cos{\tilde\varphi}\leq0.
\end{align}
So, we have
\begin{align*}
\notag\underline{P}=    P({\bf x}(\varphi))&=\frac{3}{2}\overline{I}V({\bf x}(\varphi))\cos{\varphi}\\
&<\frac{3}{2}\overline{I}V({\bf x}(\tilde\varphi))\cos{\tilde\varphi}=P({\bf x}(\tilde\varphi)).
\end{align*}

This implies that $\tilde\varphi$ is  better than $\varphi^\star$, since it  results in lower voltage without making the minimum power constraint binding. However, this contradicts (\ref{bindinggp}). Therefore, there is no better solution than $\varphi^\star$ and ${\bf x}(\varphi^\star)$ is the optimum of P2. This completes the proof of Claim 3. \qed

\emph{Claim 4:} If neither of C1 and C2 hold, P2 is an exact relaxation of P0$^\prime$. 

\emph{Proof:} Firstly, it follows that $P({\bf x}_2^\star)=P({\bf x}_3^\star)=\underline{P}<\overline{P}$. Since P1 is a relaxation of P2, we have $V({\bf x}^\star_2)\geq V({\bf x}^\star_1)>0$. This implies that both the upper bound of active power and voltage constraints are not violated at the solution of P2, which indicates that P2 is an exact relaxation of P0$^\prime$. This concludes the proof. 
\qed

\section{Transformation From $
\rm DQ^-$ to $\rm DQ^+$}
The unbalanced three-phase grid voltage and PCC voltage (with no zero-sequence) can be expressed as,
\begin{align*}
v_{g,a}(t)=&V^+_g\sin(\omega t+\theta^+_g)+V^-_g\sin(\omega t+\theta^-_g)\\
v_{g,b}(t)=&V^+_g\sin(\omega t+\theta^+_g-120^\circ)\\
&+V^-_g\sin(\omega t+\theta^-_g+120^\circ)\\
v_{g,c}(t)=&V^+_g\sin(\omega t+\theta^+_g -120^\circ)\\
&+V^-_g\sin(\omega t+\theta^-_g-120^\circ)
\end{align*}
and 
\begin{align*}
v_a(t)&=V^+\sin(\omega t+\theta^+)+V^-\sin(\omega t+\theta^-)\\
v_b(t)&=V^+\sin(\omega t+\theta^+-120^\circ)+V^-\sin(\omega t+\theta^-+120^\circ)\\
v_c(t)&=V^+\sin(\omega t+\theta^+ -120^\circ)+V^-\sin(\omega t+\theta^--120^\circ).
\end{align*}

The angle references for $\rm DQ^-$, $\rm DQ^-_g$ and $\rm DQ^+$ are:
\begin{align*}
{\rm DQ}^-&:\omega t+\theta^-\\
{\rm DQ^-_g}&:\omega t+\theta^-_g\\
{\rm DQ}^+&:-(\omega t+\theta^+)
\end{align*}

To be consistent with the modeling and analysis in Sections II--IV, let $(I_d^
\star,I_q^
\star)$ be the dq currents under $\rm DQ^-$ and $(I_d^-,I_q^-)$ be the counterpart under $\rm DQ^+$, respectively. Since the instantaneous three-phase currents are the same for different dq-frames, we have
\begin{align}
 \notag   \begin{bmatrix}
     I_d^-\\
     I_q^-
    \end{bmatrix}&={\bf T}^{\rm DQ^+}_{\rm s
    \rightarrow r}(-\alpha^+) {\bf T}^{\rm DQ^-}_{\rm r
    \rightarrow s}(\alpha^-) \begin{bmatrix}
     I_d^\star\\
     I_q^\star
    \end{bmatrix}\\ \notag &= \begin{bmatrix}
     -I^-\cos(\theta^--\theta^++\varphi^-)\\
     I^-\sin(\theta^--\theta^++\varphi^-)
    \end{bmatrix}
\end{align}
where 
$$\alpha^+=\omega t+\theta^+,\,\, \alpha^-=\omega t+\theta^-$$
$$
   {\bf T}^{\rm DQ^+}_{\rm s
    \rightarrow r}=
   \begin{bmatrix}
     -\sin(\alpha^+)&\cos(\alpha^+)\\
     -\sin(\alpha^++120^\circ)&\cos(\alpha^++120^\circ)\\
     -\sin(\alpha^+-120^\circ)    &\cos(\alpha^+-120^\circ)
    \end{bmatrix}^T
$$
$${\bf T}^{\rm DQ^-}_{\rm r
    \rightarrow s}=\begin{bmatrix}
     \sin \alpha^- &\cos\alpha^- \\
    \sin(\alpha^- +120^\circ)&\cos(\alpha^-+120^\circ)\\
    \sin(\alpha^- -120^\circ)&\cos(\alpha^- -120^\circ)
    \end{bmatrix}$$
$$I^-=\sqrt{(I_d^\star)^2+(I_q^\star)^2},\,\,\varphi^-={\rm atan2}\left({I_q^\star},{I_d^\star}\right).$$

\bibliographystyle{IEEEtran}
\bibliography{references}

\begin{thebibliography}{10}
\providecommand{\url}[1]{#1}
\csname url@samestyle\endcsname
\providecommand{\newblock}{\relax}
\providecommand{\bibinfo}[2]{#2}
\providecommand{\BIBentrySTDinterwordspacing}{\spaceskip=0pt\relax}
\providecommand{\BIBentryALTinterwordstretchfactor}{4}
\providecommand{\BIBentryALTinterwordspacing}{\spaceskip=\fontdimen2\font plus
\BIBentryALTinterwordstretchfactor\fontdimen3\font minus
  \fontdimen4\font\relax}
\providecommand{\BIBforeignlanguage}[2]{{%
\expandafter\ifx\csname l@#1\endcsname\relax
\typeout{** WARNING: IEEEtran.bst: No hyphenation pattern has been}%
\typeout{** loaded for the language `#1'. Using the pattern for}%
\typeout{** the default language instead.}%
\else
\language=\csname l@#1\endcsname
\fi
#2}}
\providecommand{\BIBdecl}{\relax}
\BIBdecl

\bibitem{VJA2001}
A.~von Jouanne and B.~Banerjee, ``Assessment of voltage unbalance,'' \emph{IEEE
  Trans. Power Del.}, vol.~16, no.~4, pp. 782--790, 2001.

\bibitem{CPT2009}
P.-T. Cheng, C.-A. Chen, T.-L. Lee, and S.-Y. Kuo, ``A cooperative imbalance
  compensation method for distributed-generation interface converters,''
  \emph{IEEE Trans. Ind. Appl.}, vol.~45, no.~2, pp. 805--815, 2009.

\bibitem{GJM2013}
J.~M. Guerrero, P.~C. Loh, T.-L. Lee, and M.~Chandorkar, ``Advanced control
  architectures for intelligent microgrids—{P}art {II}: Power quality, energy
  storage, and ac/dc microgrids,'' \emph{IEEE Trans. Ind. Electron.}, vol.~60,
  no.~4, pp. 1263--1270, 2013.

\bibitem{NT2015}
T.~Neumann, T.~Wijnhoven, G.~Deconinck, and I.~Erlich, ``Enhanced dynamic
  voltage control of type 4 wind turbines during unbalanced grid faults,''
  \emph{IEEE Trans. Energy Conv.}, vol.~30, no.~4, pp. 1650--1659, 2015.

\bibitem{CA2021}
A.~Camacho, M.~Castilla, J.~Miret, M.~Velasco, and R.~Guzman, ``Positive
  sequence voltage control, full negative sequence cancellation and current
  limitation for static compensators,'' \emph{IEEE J. Emerg. Sel. Topics Power
  Electron.}, pp. 1--1, 2021.

\bibitem{MYARI2008}
Y.~A.-R.~I. Mohamed and E.~F. El-Saadany, ``A control scheme for {PWM}
  voltage-source distributed-generation inverters for fast load-voltage
  regulation and effective mitigation of unbalanced voltage disturbances,''
  \emph{IEEE Trans. Ind. Electron.}, vol.~55, no.~5, pp. 2072--2084, 2008.

\bibitem{SM2012}
M.~Savaghebi, A.~Jalilian, J.~C. Vasquez, and J.~M. Guerrero, ``Secondary
  control scheme for voltage unbalance compensation in an islanded
  droop-controlled microgrid,'' \emph{IEEE Trans. Smart Grid}, vol.~3, no.~2,
  pp. 797--807, 2012.

\bibitem{YJ2013}
J.~Yao, H.~Li, Z.~Chen, X.~Xia, X.~Chen, Q.~Li, and Y.~Liao, ``Enhanced control
  of a {DFIG}-based wind-power generation system with series grid-side
  converter under unbalanced grid voltage conditions,'' \emph{IEEE Trans. Power
  Electron.}, vol.~28, no.~7, pp. 3167--3181, 2013.

\bibitem{LTL2013}
T.-L. Lee, S.-H. Hu, and Y.-H. Chan, ``D-{STATCOM} with positive-sequence
  admittance and negative-sequence conductance to mitigate voltage fluctuations
  in high-level penetration of distributed-generation systems,'' \emph{IEEE
  Trans. Ind. Electron.}, vol.~60, no.~4, pp. 1417--1428, 2013.

\bibitem{GF2015}
F.~Guo, C.~Wen, J.~Mao, J.~Chen, and Y.-D. Song, ``Distributed cooperative
  secondary control for voltage unbalance compensation in an islanded
  microgrid,'' \emph{IEEE Trans. Ind. Informat.}, vol.~11, no.~5, pp.
  1078--1088, 2015.

\bibitem{MNR2017}
N.~R. Merritt, C.~Chakraborty, and P.~Bajpai, ``New voltage control strategies
  for {VSC}-based {DG} units in an unbalanced microgrid,'' \emph{IEEE Trans.
  Sustain. Energy}, vol.~8, no.~3, pp. 1127--1139, 2017.

\bibitem{AS2019}
S.~Acharya, M.~S. El-Moursi, A.~Al-Hinai, A.~S. Al-Sumaiti, and H.~H.
  Zeineldin, ``A control strategy for voltage unbalance mitigation in an
  islanded microgrid considering demand side management capability,''
  \emph{IEEE Trans. Smart Grid}, vol.~10, no.~3, pp. 2558--2568, 2019.

\bibitem{GMM2020}
M.~M. Ghahderijani, A.~Camacho, C.~Moreira, M.~Castilla, and L.~García~de
  Vicuña, ``Imbalance-voltage mitigation in an inverter-based distributed
  generation system using a minimum current-based control strategy,''
  \emph{IEEE Trans. Power Del.}, vol.~35, no.~3, pp. 1399--1409, 2020.

\bibitem{BAM2021}
{\'A}.~Borrell, M.~Velasco, J.~Miret, A.~Camacho, P.~Martí, and M.~Castilla,
  ``Collaborative voltage unbalance elimination in grid-connected ac microgrids
  with grid-feeding inverters,'' \emph{IEEE Trans. Power Electron.}, vol.~36,
  no.~6, pp. 7189--7201, 2021.

\bibitem{YG2021}
\BIBentryALTinterwordspacing
Y.~Guo, B.~C. Pal, and R.~A. Jabr, ``Global optimality of inverter dynamic
  voltage support,'' \emph{arXiv:2106.16096}, 2021. [Online]. Available:
  \url{https://arxiv.org/abs/2106.16096}
\BIBentrySTDinterwordspacing

\bibitem{LK2007}
K.~Li, J.~Liu, Z.~Wang, and B.~Wei, ``Strategies and operating point
  optimization of statcom control for voltage unbalance mitigation in
  three-phase three-wire systems,'' \emph{IEEE Trans. Power Del.}, vol.~22,
  no.~1, pp. 413--422, 2007.

\bibitem{NF2016}
F.~Nejabatkhah, Y.~W. Li, and B.~Wu, ``Control strategies of three-phase
  distributed generation inverters for grid unbalanced voltage compensation,''
  \emph{IEEE Trans. Power Electron.}, vol.~31, no.~7, pp. 5228--5241, 2016.

\bibitem{CA2018}
A.~Camacho, M.~Castilla, J.~Miret, L.~G. de~Vicuña, and R.~Guzman, ``Positive
  and negative sequence control strategies to maximize the voltage support in
  resistive–inductive grids during grid faults,'' \emph{IEEE Trans. Power
  Electron.}, vol.~33, no.~6, pp. 5362--5373, 2018.

\bibitem{SMA2019}
M.~A. Shuvra and B.~Chowdhury, ``Distributed dynamic grid support using smart
  {PV} inverters during unbalanced grid faults,'' \emph{IET Renew. Power
  Gener.}, vol.~13, no.~4, pp. 598--608, 2019.

\bibitem{PP2001}
P.~Pillay and M.~Manyage, ``Definitions of voltage unbalance,'' \emph{IEEE
  Power Eng. Rev.}, vol.~21, no.~5, pp. 49--51, May 2001.

\bibitem{GO2014}
{\"O}.~Göksu, R.~Teodorescu, C.~L. Bak, F.~Iov, and P.~C. Kjær, ``Instability
  of wind turbine converters during current injection to low voltage grid
  faults and {PLL} frequency based stability solution,'' \emph{IEEE Trans.
  Power Syst.}, vol.~29, no.~4, pp. 1683--1691, 2014.

\bibitem{GH2018}
H.~Geng, L.~Liu, and R.~Li, ``Synchronization and reactive current support of
  {PMSG}-based wind farm during severe grid fault,'' \emph{IEEE Trans. Sustain.
  Energy}, vol.~9, no.~4, pp. 1596--1604, 2018.

\bibitem{WB2015}
B.~Weise, ``Impact of k-factor and active current reduction during
  fault-ride-through of generating units connected via voltage-sourced
  converters on power system stability,'' \emph{IET Renew. Power Gener.},
  vol.~9, no.~1, pp. 25--36, 2015.

\bibitem{YA2006}
A.~Yazdani and R.~Iravani, ``A unified dynamic model and control for the
  voltage-sourced converter under unbalanced grid conditions,'' \emph{IEEE
  Trans. Power Del.}, vol.~21, no.~3, pp. 1620--1629, 2006.

\bibitem{TMG2020}
M.~G. Taul, R.~E. Betz, and F.~Blaabjerg, ``Rapid impedance estimation
  algorithm for mitigation of synchronization instability of paralleled
  converters under grid faults,'' in \emph{Proc. IEEE 22nd Eur. Conf. Power
  Electron. Appl.}, Sep. 2020, pp. 1--5.

\bibitem{MN2021}
N.~Mohammed, T.~Kerekes, and M.~Ciobotaru, ``{An online event-based grid
  impedance estimation technique using grid-connected inverters},'' \emph{IEEE
  Trans. Power Electron.}, vol.~36, no.~5, pp. 6106--6117, 2021.

\end{thebibliography}

\end{document}